\documentclass[longauth]{aa}

\usepackage{graphicx}
\usepackage{natbib}
\usepackage{scalerel}

\usepackage[table]{xcolor}

\usepackage{txfonts}
\usepackage[pdfencoding=auto,psdextra]{hyperref}
\hypersetup{
    colorlinks=true,
    linkcolor=blue,
    filecolor=magenta,      
    urlcolor=blue,
    citecolor=blue
}
\makeatletter
\renewcommand*\aa@pageof{, page \thepage{} of \pageref*{LastPage}}
\makeatother

\usepackage[utf8]{inputenc}

\usepackage[switch, modulo]{lineno}
              
\renewcommand{\linenumbers}[0]{}

\usepackage{euclid}

\begin{document}

\title{\Euclid: Data Release 1 (DR1) -- Fornax-7, an ultra-faint companion to the Fornax dwarf spheroidal galaxy?\thanks{This paper is published on
behalf of the Euclid Consortium}}

\subtitle{A remote star cluster or dwarf satellite of the Fornax dSph}

\newcommand{\orcid}[1]{} 
\author{T.~Saifollahi\orcid{0000-0002-9554-7660}\thanks{\email{Teymoor.saifollahi@astro.unistra.fr}}\inst{\ref{aff1},\ref{aff2}}
\and A.~Lan\c{c}on\orcid{0000-0002-7214-8296}\inst{\ref{aff1}}
\and A.~M.~N.~Ferguson\orcid{0000-0001-7934-1278}\inst{\ref{aff3}}
\and P.~Boldrini\inst{\ref{aff4}}
\and M.~Gatto\orcid{0000-0003-4636-6457}\inst{\ref{aff5}}
\and J.~M.~Howell\orcid{0009-0002-2242-6515}\inst{\ref{aff3}}
\and N.~F.~Martin\orcid{0000-0002-1349-202X}\inst{\ref{aff1},\ref{aff6}}
\and A.~C.~Robin\orcid{0000-0001-8654-9499}\inst{\ref{aff7}}
\and F.~Annibali\inst{\ref{aff8}}
\and M.~Baes\orcid{0000-0002-3930-2757}\inst{\ref{aff9}}
\and E.~Balbinot\orcid{0000-0002-1322-3153}\inst{\ref{aff10},\ref{aff11}}
\and G.~Battaglia\orcid{0000-0002-6551-4294}\inst{\ref{aff12},\ref{aff13}}
\and M.~Bellazzini\orcid{0000-0001-8200-810X}\inst{\ref{aff8}}
\and Michele~Cantiello\orcid{0000-0003-2072-384X}\inst{\ref{aff14}}
\and M.-R.~L.~Cioni\orcid{0000-0002-6797-696X}\inst{\ref{aff15}}
\and C.~Crociati\orcid{0009-0002-8571-5170}\inst{\ref{aff3}}
\and V.~Dornan\orcid{0000-0002-7731-1291}\inst{\ref{aff3}}
\and P.-A.~Duc\orcid{0000-0003-3343-6284}\inst{\ref{aff1}}
\and A.~Genina\orcid{0000-0003-0073-3012}\inst{\ref{aff3}}
\and C.~M.~Gutierrez\orcid{0000-0001-7854-783X}\inst{\ref{aff12},\ref{aff13}}
\and R.~Habas\orcid{0000-0002-4033-3841}\inst{\ref{aff14}}
\and P.~Jablonka\orcid{0000-0002-9655-1063}\inst{\ref{aff16}}
\and S.~S.~Larsen\orcid{0000-0003-0069-1203}\inst{\ref{aff17}}
\and M.~Libralato\orcid{0000-0001-9673-7397}\inst{\ref{aff18}}
\and F.~R.~Marleau\orcid{0000-0002-1442-2947}\inst{\ref{aff19}}
\and D.~Mart\'inez-Delgado\inst{\ref{aff20},\ref{aff21}}
\and S.~Martocchia\orcid{0000-0001-7110-6775}\inst{\ref{aff22}}
\and D.~Massari\orcid{0000-0001-8892-4301}\inst{\ref{aff8}}
\and F.~Niederhofer\orcid{0000-0002-4341-9819}\inst{\ref{aff15}}
\and J.~D.~Sakowska\orcid{0000-0002-1594-1466}\inst{\ref{aff23}}
\and M.~Urbano\orcid{0000-0001-5640-0650}\inst{\ref{aff24},\ref{aff1}}
\and L.~R.~Bedin\orcid{0000-0003-4080-6466}\inst{\ref{aff18}}
\and M.~Griggio\orcid{0000-0002-5060-1379}\inst{\ref{aff25}}
\and A.~Mohandasan\orcid{0000-0001-5182-0330}\inst{\ref{aff7}}
\and B.~Altieri\orcid{0000-0003-3936-0284}\inst{\ref{aff26}}
\and S.~Andreon\orcid{0000-0002-2041-8784}\inst{\ref{aff27}}
\and N.~Auricchio\orcid{0000-0003-4444-8651}\inst{\ref{aff8}}
\and C.~Baccigalupi\orcid{0000-0002-8211-1630}\inst{\ref{aff28},\ref{aff29},\ref{aff30},\ref{aff31}}
\and M.~Baldi\orcid{0000-0003-4145-1943}\inst{\ref{aff32},\ref{aff8},\ref{aff33}}
\and S.~Bardelli\orcid{0000-0002-8900-0298}\inst{\ref{aff8}}
\and P.~Battaglia\orcid{0000-0002-7337-5909}\inst{\ref{aff8}}
\and A.~Biviano\orcid{0000-0002-0857-0732}\inst{\ref{aff29},\ref{aff28}}
\and M.~Bolzonella\orcid{0000-0003-3278-4607}\inst{\ref{aff8}}
\and E.~Branchini\orcid{0000-0002-0808-6908}\inst{\ref{aff34},\ref{aff35},\ref{aff27}}
\and M.~Brescia\orcid{0000-0001-9506-5680}\inst{\ref{aff36},\ref{aff5}}
\and S.~Camera\orcid{0000-0003-3399-3574}\inst{\ref{aff37},\ref{aff38},\ref{aff39}}
\and V.~Capobianco\orcid{0000-0002-3309-7692}\inst{\ref{aff39}}
\and C.~Carbone\orcid{0000-0003-0125-3563}\inst{\ref{aff40}}
\and J.~Carretero\orcid{0000-0002-3130-0204}\inst{\ref{aff41},\ref{aff42}}
\and M.~Castellano\orcid{0000-0001-9875-8263}\inst{\ref{aff43}}
\and G.~Castignani\orcid{0000-0001-6831-0687}\inst{\ref{aff8}}
\and S.~Cavuoti\orcid{0000-0002-3787-4196}\inst{\ref{aff5},\ref{aff44}}
\and K.~C.~Chambers\orcid{0000-0001-6965-7789}\inst{\ref{aff45}}
\and A.~Cimatti\inst{\ref{aff46}}
\and C.~Colodro-Conde\inst{\ref{aff12}}
\and G.~Congedo\orcid{0000-0003-2508-0046}\inst{\ref{aff3}}
\and C.~J.~Conselice\orcid{0000-0003-1949-7638}\inst{\ref{aff47}}
\and L.~Conversi\orcid{0000-0002-6710-8476}\inst{\ref{aff48},\ref{aff26}}
\and Y.~Copin\orcid{0000-0002-5317-7518}\inst{\ref{aff49}}
\and A.~Costille\inst{\ref{aff22}}
\and F.~Courbin\orcid{0000-0003-0758-6510}\inst{\ref{aff50},\ref{aff51},\ref{aff52}}
\and H.~M.~Courtois\orcid{0000-0003-0509-1776}\inst{\ref{aff53}}
\and M.~Cropper\orcid{0000-0003-4571-9468}\inst{\ref{aff54}}
\and H.~Degaudenzi\orcid{0000-0002-5887-6799}\inst{\ref{aff55}}
\and G.~De~Lucia\orcid{0000-0002-6220-9104}\inst{\ref{aff29}}
\and H.~Dole\orcid{0000-0002-9767-3839}\inst{\ref{aff56}}
\and F.~Dubath\orcid{0000-0002-6533-2810}\inst{\ref{aff55}}
\and X.~Dupac\inst{\ref{aff26}}
\and M.~Farina\orcid{0000-0002-3089-7846}\inst{\ref{aff57}}
\and R.~Farinelli\inst{\ref{aff8}}
\and F.~Faustini\orcid{0000-0001-6274-5145}\inst{\ref{aff43}}
\and S.~Ferriol\inst{\ref{aff49}}
\and M.~Frailis\orcid{0000-0002-7400-2135}\inst{\ref{aff29}}
\and E.~Franceschi\orcid{0000-0002-0585-6591}\inst{\ref{aff8}}
\and M.~Fumana\orcid{0000-0001-6787-5950}\inst{\ref{aff40}}
\and L.~Gabarra\orcid{0000-0002-8486-8856}\inst{\ref{aff58}}
\and S.~Galeotta\orcid{0000-0002-3748-5115}\inst{\ref{aff29}}
\and K.~George\orcid{0000-0002-1734-8455}\inst{\ref{aff59}}
\and B.~Gillis\orcid{0000-0002-4478-1270}\inst{\ref{aff3}}
\and C.~Giocoli\orcid{0000-0002-9590-7961}\inst{\ref{aff8},\ref{aff33}}
\and J.~Gracia-Carpio\orcid{0000-0003-4689-3134}\inst{\ref{aff60}}
\and A.~Grazian\orcid{0000-0002-5688-0663}\inst{\ref{aff18}}
\and F.~Grupp\inst{\ref{aff60},\ref{aff61}}
\and S.~V.~H.~Haugan\orcid{0000-0001-9648-7260}\inst{\ref{aff62}}
\and H.~Hoekstra\orcid{0000-0002-0641-3231}\inst{\ref{aff11}}
\and W.~Holmes\orcid{0009-0007-8554-4646}\inst{\ref{aff63}}
\and I.~M.~Hook\orcid{0000-0002-2960-978X}\inst{\ref{aff64}}
\and F.~Hormuth\inst{\ref{aff65}}
\and A.~Hornstrup\orcid{0000-0002-3363-0936}\inst{\ref{aff66},\ref{aff67}}
\and K.~Jahnke\orcid{0000-0003-3804-2137}\inst{\ref{aff6}}
\and M.~Jhabvala\inst{\ref{aff68}}
\and S.~Kermiche\orcid{0000-0002-0302-5735}\inst{\ref{aff69}}
\and A.~Kiessling\orcid{0000-0002-2590-1273}\inst{\ref{aff63}}
\and B.~Kubik\orcid{0009-0006-5823-4880}\inst{\ref{aff49}}
\and K.~Kuijken\orcid{0000-0002-3827-0175}\inst{\ref{aff11}}
\and M.~K\"ummel\orcid{0000-0003-2791-2117}\inst{\ref{aff61}}
\and M.~Kunz\orcid{0000-0002-3052-7394}\inst{\ref{aff70}}
\and H.~Kurki-Suonio\orcid{0000-0002-4618-3063}\inst{\ref{aff71},\ref{aff72}}
\and A.~M.~C.~Le~Brun\orcid{0000-0002-0936-4594}\inst{\ref{aff73}}
\and S.~Ligori\orcid{0000-0003-4172-4606}\inst{\ref{aff39}}
\and P.~B.~Lilje\orcid{0000-0003-4324-7794}\inst{\ref{aff62}}
\and V.~Lindholm\orcid{0000-0003-2317-5471}\inst{\ref{aff71},\ref{aff72}}
\and I.~Lloro\orcid{0000-0001-5966-1434}\inst{\ref{aff74}}
\and M.~Magliocchetti\orcid{0000-0001-9158-4838}\inst{\ref{aff57}}
\and G.~Mainetti\orcid{0000-0003-2384-2377}\inst{\ref{aff75}}
\and O.~Mansutti\orcid{0000-0001-5758-4658}\inst{\ref{aff29}}
\and O.~Marggraf\orcid{0000-0001-7242-3852}\inst{\ref{aff76}}
\and M.~Martinelli\orcid{0000-0002-6943-7732}\inst{\ref{aff43},\ref{aff77}}
\and N.~Martinet\orcid{0000-0003-2786-7790}\inst{\ref{aff22}}
\and F.~Marulli\orcid{0000-0002-8850-0303}\inst{\ref{aff78},\ref{aff8},\ref{aff33}}
\and R.~J.~Massey\orcid{0000-0002-6085-3780}\inst{\ref{aff79}}
\and E.~Medinaceli\orcid{0000-0002-4040-7783}\inst{\ref{aff8}}
\and M.~Meneghetti\orcid{0000-0003-1225-7084}\inst{\ref{aff8},\ref{aff33}}
\and E.~Merlin\orcid{0000-0001-6870-8900}\inst{\ref{aff18}}
\and G.~Meylan\orcid{0000-0001-6503-0209}\inst{\ref{aff16}}
\and P.~Monaco\orcid{0000-0003-2083-7564}\inst{\ref{aff80},\ref{aff29},\ref{aff30},\ref{aff28}}
\and A.~Mora\orcid{0000-0002-1922-8529}\inst{\ref{aff81}}
\and M.~Moresco\orcid{0000-0002-7616-7136}\inst{\ref{aff78},\ref{aff8}}
\and C.~Moretti\orcid{0000-0003-3314-8936}\inst{\ref{aff29},\ref{aff28},\ref{aff30}}
\and L.~Moscardini\orcid{0000-0002-3473-6716}\inst{\ref{aff78},\ref{aff8},\ref{aff33}}
\and R.~Nakajima\orcid{0009-0009-1213-7040}\inst{\ref{aff76}}
\and C.~Neissner\orcid{0000-0001-8524-4968}\inst{\ref{aff82},\ref{aff42}}
\and S.-M.~Niemi\orcid{0009-0005-0247-0086}\inst{\ref{aff83}}
\and C.~Padilla\orcid{0000-0001-7951-0166}\inst{\ref{aff82}}
\and S.~Paltani\orcid{0000-0002-8108-9179}\inst{\ref{aff55}}
\and F.~Pasian\orcid{0000-0002-4869-3227}\inst{\ref{aff29}}
\and W.~J.~Percival\orcid{0000-0002-0644-5727}\inst{\ref{aff84},\ref{aff85},\ref{aff86}}
\and V.~Pettorino\orcid{0000-0002-4203-9320}\inst{\ref{aff83}}
\and A.~Pezzotta\orcid{0000-0003-0726-2268}\inst{\ref{aff27}}
\and G.~Polenta\orcid{0000-0003-4067-9196}\inst{\ref{aff87}}
\and M.~Poncet\inst{\ref{aff2}}
\and L.~A.~Popa\inst{\ref{aff88}}
\and F.~Raison\orcid{0000-0002-7819-6918}\inst{\ref{aff60}}
\and A.~Renzi\orcid{0000-0001-9856-1970}\inst{\ref{aff89},\ref{aff90},\ref{aff8}}
\and J.~Rhodes\orcid{0000-0002-4485-8549}\inst{\ref{aff63}}
\and G.~Riccio\inst{\ref{aff5}}
\and I.~Risso\orcid{0000-0003-2525-7761}\inst{\ref{aff34},\ref{aff35},\ref{aff27}}
\and E.~Romelli\orcid{0000-0003-3069-9222}\inst{\ref{aff29}}
\and M.~Roncarelli\orcid{0000-0001-9587-7822}\inst{\ref{aff8}}
\and B.~Rusholme\orcid{0000-0001-7648-4142}\inst{\ref{aff91}}
\and R.~Saglia\orcid{0000-0003-0378-7032}\inst{\ref{aff61},\ref{aff60}}
\and Z.~Sakr\orcid{0000-0002-4823-3757}\inst{\ref{aff92},\ref{aff93},\ref{aff94}}
\and D.~Sapone\orcid{0000-0001-7089-4503}\inst{\ref{aff95}}
\and M.~Schirmer\orcid{0000-0003-2568-9994}\inst{\ref{aff6}}
\and P.~Schneider\orcid{0000-0001-8561-2679}\inst{\ref{aff76}}
\and A.~Secroun\orcid{0000-0003-0505-3710}\inst{\ref{aff69}}
\and E.~Sihvola\orcid{0000-0003-1804-7715}\inst{\ref{aff96}}
\and C.~Sirignano\orcid{0000-0002-0995-7146}\inst{\ref{aff89},\ref{aff90}}
\and G.~Sirri\orcid{0000-0003-2626-2853}\inst{\ref{aff33}}
\and L.~Stanco\orcid{0000-0002-9706-5104}\inst{\ref{aff90}}
\and P.~Tallada-Cresp\'{i}\orcid{0000-0002-1336-8328}\inst{\ref{aff41},\ref{aff42}}
\and A.~N.~Taylor\inst{\ref{aff3}}
\and I.~Tereno\orcid{0000-0002-4537-6218}\inst{\ref{aff97},\ref{aff98}}
\and S.~Toft\orcid{0000-0003-3631-7176}\inst{\ref{aff99},\ref{aff100}}
\and R.~Toledo-Moreo\orcid{0000-0002-2997-4859}\inst{\ref{aff101},\ref{aff102}}
\and F.~Torradeflot\orcid{0000-0003-1160-1517}\inst{\ref{aff42},\ref{aff41}}
\and A.~Tsyganov\inst{\ref{aff103}}
\and I.~Tutusaus\orcid{0000-0002-3199-0399}\inst{\ref{aff104},\ref{aff105},\ref{aff93}}
\and J.~Valiviita\orcid{0000-0001-6225-3693}\inst{\ref{aff71},\ref{aff72}}
\and T.~Vassallo\orcid{0000-0001-6512-6358}\inst{\ref{aff29},\ref{aff59}}
\and Y.~Wang\orcid{0000-0002-4749-2984}\inst{\ref{aff91}}
\and J.~Weller\orcid{0000-0002-8282-2010}\inst{\ref{aff61},\ref{aff60}}
\and A.~Zacchei\orcid{0000-0003-0396-1192}\inst{\ref{aff29},\ref{aff28}}
\and F.~M.~Zerbi\orcid{0000-0002-9996-973X}\inst{\ref{aff27}}
\and I.~A.~Zinchenko\orcid{0000-0002-2944-2449}\inst{\ref{aff106}}
\and E.~Zucca\orcid{0000-0002-5845-8132}\inst{\ref{aff8}}
\and J.~Garc\'ia-Bellido\orcid{0000-0002-9370-8360}\inst{\ref{aff92}}
\and K.~Tanidis\orcid{0000-0001-9843-5130}\inst{\ref{aff107}}}
										   
\institute{Universit\'e de Strasbourg, CNRS, Observatoire astronomique de Strasbourg, UMR 7550, 67000 Strasbourg, France\label{aff1}
\and
Centre National d'Etudes Spatiales -- Centre spatial de Toulouse, 18 avenue Edouard Belin, 31401 Toulouse Cedex 9, France\label{aff2}
\and
Institute for Astronomy, University of Edinburgh, Royal Observatory, Blackford Hill, Edinburgh EH9 3HJ, UK\label{aff3}
\and
Observatoire de Paris, PSL Research University 61, avenue de l'Observatoire, 75014 Paris, France\label{aff4}
\and
INAF-Osservatorio Astronomico di Capodimonte, Via Moiariello 16, 80131 Napoli, Italy\label{aff5}
\and
Max-Planck-Institut f\"ur Astronomie, K\"onigstuhl 17, 69117 Heidelberg, Germany\label{aff6}
\and
Universite Marie et Louis Pasteur, CNRS, Observatoire des Sciences de l'Univers THETA Franche-Comte Bourgogne, Institut UTINAM, Observatoire de Besan\c con, BP 1615, 25010 Besan\c con Cedex, France\label{aff7}
\and
INAF-Osservatorio di Astrofisica e Scienza dello Spazio di Bologna, Via Piero Gobetti 93/3, 40129 Bologna, Italy\label{aff8}
\and
Universiteit Gent, Department of Physics and Astronomy, Proeftuinstraat 86 N3, 9000 Ghent, Belgium
\label{aff9}
\and
Kapteyn Astronomical Institute, University of Groningen, PO Box 800, 9700 AV Groningen, The Netherlands\label{aff10}
\and
Leiden Observatory, Leiden University, Einsteinweg 55, 2333 CC Leiden, The Netherlands\label{aff11}
\and
Instituto de Astrof\'{\i}sica de Canarias, E-38205 La Laguna, Tenerife, Spain\label{aff12}
\and
Universidad de La Laguna, Dpto. Astrof\'\i sica, E-38206 La Laguna, Tenerife, Spain\label{aff13}
\and
INAF - Osservatorio Astronomico d'Abruzzo, Via Maggini, 64100, Teramo, Italy\label{aff14}
\and
Leibniz-Institut f\"{u}r Astrophysik (AIP), An der Sternwarte 16, 14482 Potsdam, Germany\label{aff15}
\and
Institute of Physics, Laboratory of Astrophysics, Ecole Polytechnique F\'ed\'erale de Lausanne (EPFL), Observatoire de Sauverny, 1290 Versoix, Switzerland\label{aff16}
\and
Department of Astrophysics/IMAPP, Radboud University, PO Box 9010, 6500 GL Nijmegen, The Netherlands\label{aff17}
\and
INAF-Osservatorio Astronomico di Padova, Via dell'Osservatorio 5, 35122 Padova, Italy\label{aff18}
\and
Universit\"at Innsbruck, Institut f\"ur Astro- und Teilchenphysik, Technikerstr. 25/8, 6020 Innsbruck, Austria\label{aff19}
\and
Centro de Estudios de F\'isica del Cosmos de Arag\'on (CEFCA), Plaza San Juan, 1, planta 2, 44001, Teruel, Spain\label{aff20}
\and
ARAID Foundation, Avda. de Ranillas, 1-D, E-50018 Zaragoza, Spain\label{aff21}
\and
Aix-Marseille Universit\'e, CNRS, CNES, LAM, Marseille, France\label{aff22}
\and
Instituto de Astrof\'isica de Andaluc\'ia, CSIC, Glorieta de la Astronom\'\i a, 18080, Granada, Spain\label{aff23}
\and
School of Physics and Astronomy, University of Nottingham, University Park, Nottingham NG7 2RD, UK\label{aff24}
\and
Space Telescope Science Institute, 3700 San Martin Dr, Baltimore, MD 21218, USA\label{aff25}
\and
ESAC/ESA, Camino Bajo del Castillo, s/n., Urb. Villafranca del Castillo, 28692 Villanueva de la Ca\~nada, Madrid, Spain\label{aff26}
\and
INAF-Osservatorio Astronomico di Brera, Via Brera 28, 20122 Milano, Italy\label{aff27}
\and
IFPU, Institute for Fundamental Physics of the Universe, via Beirut 2, 34151 Trieste, Italy\label{aff28}
\and
INAF-Osservatorio Astronomico di Trieste, Via G. B. Tiepolo 11, 34143 Trieste, Italy\label{aff29}
\and
INFN, Sezione di Trieste, Via Valerio 2, 34127 Trieste TS, Italy\label{aff30}
\and
SISSA, International School for Advanced Studies, Via Bonomea 265, 34136 Trieste TS, Italy\label{aff31}
\and
Dipartimento di Fisica e Astronomia, Universit\`a di Bologna, Via Gobetti 93/2, 40129 Bologna, Italy\label{aff32}
\and
INFN-Sezione di Bologna, Viale Berti Pichat 6/2, 40127 Bologna, Italy\label{aff33}
\and
Dipartimento di Fisica, Universit\`a di Genova, Via Dodecaneso 33, 16146, Genova, Italy\label{aff34}
\and
INFN-Sezione di Genova, Via Dodecaneso 33, 16146, Genova, Italy\label{aff35}
\and
Department of Physics "E. Pancini", University Federico II, Via Cinthia 6, 80126, Napoli, Italy\label{aff36}
\and
Dipartimento di Fisica, Universit\`a degli Studi di Torino, Via P. Giuria 1, 10125 Torino, Italy\label{aff37}
\and
INFN-Sezione di Torino, Via P. Giuria 1, 10125 Torino, Italy\label{aff38}
\and
INAF-Osservatorio Astrofisico di Torino, Via Osservatorio 20, 10025 Pino Torinese (TO), Italy\label{aff39}
\and
INAF-IASF Milano, Via Alfonso Corti 12, 20133 Milano, Italy\label{aff40}
\and
Centro de Investigaciones Energ\'eticas, Medioambientales y Tecnol\'ogicas (CIEMAT), Avenida Complutense 40, 28040 Madrid, Spain\label{aff41}
\and
Port d'Informaci\'{o} Cient\'{i}fica, Campus UAB, C. Albareda s/n, 08193 Bellaterra (Barcelona), Spain\label{aff42}
\and
INAF-Osservatorio Astronomico di Roma, Via Frascati 33, 00078 Monteporzio Catone, Italy\label{aff43}
\and
INFN -- Sezione di Napoli, Via Cinthia 6, 80126, Napoli, Italy\label{aff44}
\and
Institute for Astronomy, University of Hawaii, 2680 Woodlawn Drive, Honolulu, HI 96822, USA\label{aff45}
\and
Dipartimento di Fisica e Astronomia "Augusto Righi" - Alma Mater Studiorum Universit\`a di Bologna, Viale Berti Pichat 6/2, 40127 Bologna, Italy\label{aff46}
\and
Jodrell Bank Centre for Astrophysics, Department of Physics and Astronomy, University of Manchester, Oxford Road, Manchester M13 9PL, UK\label{aff47}
\and
European Space Agency/ESRIN, Largo Galileo Galilei 1, 00044 Frascati, Roma, Italy\label{aff48}
\and
Universit\'e Claude Bernard Lyon 1, CNRS/IN2P3, IP2I Lyon, UMR 5822, Villeurbanne, F-69100, France\label{aff49}
\and
Institut de Ci\`{e}ncies del Cosmos (ICCUB), Universitat de Barcelona (IEEC-UB), Mart\'{i} i Franqu\`{e}s 1, 08028 Barcelona, Spain\label{aff50}
\and
Instituci\'o Catalana de Recerca i Estudis Avan\c{c}ats (ICREA), Passeig de Llu\'{\i}s Companys 23, 08010 Barcelona, Spain\label{aff51}
\and
Institut de Ciencies de l'Espai (IEEC-CSIC), Campus UAB, Carrer de Can Magrans, s/n Cerdanyola del Vall\'es, 08193 Barcelona, Spain\label{aff52}
\and
UCB Lyon 1, CNRS/IN2P3, IUF, IP2I Lyon, 4 rue Enrico Fermi, 69622 Villeurbanne, France\label{aff53}
\and
Mullard Space Science Laboratory, University College London, Holmbury St Mary, Dorking, Surrey RH5 6NT, UK\label{aff54}
\and
Department of Astronomy, University of Geneva, ch. d'Ecogia 16, 1290 Versoix, Switzerland\label{aff55}
\and
Universit\'e Paris-Saclay, CNRS, Institut d'astrophysique spatiale, 91405, Orsay, France\label{aff56}
\and
INAF-Istituto di Astrofisica e Planetologia Spaziali, via del Fosso del Cavaliere, 100, 00100 Roma, Italy\label{aff57}
\and
Department of Physics, University of Oxford, Keble Road, Oxford OX1 3RH, UK\label{aff58}
\and
University Observatory, LMU Faculty of Physics, Scheinerstr.~1, 81679 Munich, Germany\label{aff59}
\and
Max Planck Institute for Extraterrestrial Physics, Giessenbachstr. 1, 85748 Garching, Germany\label{aff60}
\and
Universit\"ats-Sternwarte M\"unchen, Fakult\"at f\"ur Physik, Ludwig-Maximilians-Universit\"at M\"unchen, Scheinerstr.~1, 81679 M\"unchen, Germany\label{aff61}
\and
Institute of Theoretical Astrophysics, University of Oslo, P.O. Box 1029 Blindern, 0315 Oslo, Norway\label{aff62}
\and
Jet Propulsion Laboratory, California Institute of Technology, 4800 Oak Grove Drive, Pasadena, CA, 91109, USA\label{aff63}
\and
Department of Physics, Lancaster University, Lancaster, LA1 4YB, UK\label{aff64}
\and
Felix Hormuth Engineering, Goethestr. 17, 69181 Leimen, Germany\label{aff65}
\and
Technical University of Denmark, Elektrovej 327, 2800 Kgs. Lyngby, Denmark\label{aff66}
\and
Cosmic Dawn Center (DAWN), Denmark\label{aff67}
\and
NASA Goddard Space Flight Center, Greenbelt, MD 20771, USA\label{aff68}
\and
Aix-Marseille Universit\'e, CNRS/IN2P3, CPPM, Marseille, France\label{aff69}
\and
Universit\'e de Gen\`eve, D\'epartement de Physique Th\'eorique and Centre for Astroparticle Physics, 24 quai Ernest-Ansermet, CH-1211 Gen\`eve 4, Switzerland\label{aff70}
\and
Department of Physics, P.O. Box 64, University of Helsinki, 00014 Helsinki, Finland\label{aff71}
\and
Helsinki Institute of Physics, Gustaf H{\"a}llstr{\"o}min katu 2, University of Helsinki, 00014 Helsinki, Finland\label{aff72}
\and
Laboratoire d'etude de l'Univers et des phenomenes eXtremes, Observatoire de Paris, Universit\'e PSL, Sorbonne Universit\'e, CNRS, 92190 Meudon, France\label{aff73}
\and
SKAO, Jodrell Bank, Lower Withington, Macclesfield SK11 9FT, UK\label{aff74}
\and
Centre de Calcul de l'IN2P3/CNRS, 21 avenue Pierre de Coubertin 69627 Villeurbanne Cedex, France\label{aff75}
\and
Universit\"at Bonn, Argelander-Institut f\"ur Astronomie, Auf dem H\"ugel 71, 53121 Bonn, Germany\label{aff76}
\and
INFN-Sezione di Roma, Piazzale Aldo Moro, 2 - c/o Dipartimento di Fisica, Edificio G. Marconi, 00185 Roma, Italy\label{aff77}
\and
Dipartimento di Fisica e Astronomia "Augusto Righi" - Alma Mater Studiorum Universit\`a di Bologna, via Piero Gobetti 93/2, 40129 Bologna, Italy\label{aff78}
\and
Department of Physics, Institute for Computational Cosmology, Durham University, South Road, Durham, DH1 3LE, UK\label{aff79}
\and
Dipartimento di Fisica - Sezione di Astronomia, Universit\`a di Trieste, Via Tiepolo 11, 34131 Trieste, Italy\label{aff80}
\and
Telespazio UK S.L. for European Space Agency (ESA), Camino bajo del Castillo, s/n, Urbanizacion Villafranca del Castillo, Villanueva de la Ca\~nada, 28692 Madrid, Spain\label{aff81}
\and
Institut de F\'{i}sica d'Altes Energies (IFAE), The Barcelona Institute of Science and Technology, Campus UAB, 08193 Bellaterra (Barcelona), Spain\label{aff82}
\and
European Space Agency/ESTEC, Keplerlaan 1, 2201 AZ Noordwijk, The Netherlands\label{aff83}
\and
Waterloo Centre for Astrophysics, University of Waterloo, Waterloo, Ontario N2L 3G1, Canada\label{aff84}
\and
Department of Physics and Astronomy, University of Waterloo, Waterloo, Ontario N2L 3G1, Canada\label{aff85}
\and
Perimeter Institute for Theoretical Physics, Waterloo, Ontario N2L 2Y5, Canada\label{aff86}
\and
Space Science Data Center, Italian Space Agency, via del Politecnico snc, 00133 Roma, Italy\label{aff87}
\and
Institute of Space Science, Str. Atomistilor, nr. 409 M\u{a}gurele, Ilfov, 077125, Romania\label{aff88}
\and
Dipartimento di Fisica e Astronomia "G. Galilei", Universit\`a di Padova, Via Marzolo 8, 35131 Padova, Italy\label{aff89}
\and
INFN-Padova, Via Marzolo 8, 35131 Padova, Italy\label{aff90}
\and
Caltech/IPAC, 1200 E. California Blvd., Pasadena, CA 91125, USA\label{aff91}
\and
Instituto de F\'isica Te\'orica UAM-CSIC, Campus de Cantoblanco, 28049 Madrid, Spain\label{aff92}
\and
Institut de Recherche en Astrophysique et Plan\'etologie (IRAP), Universit\'e de Toulouse, CNRS, UPS, CNES, 14 Av. Edouard Belin, 31400 Toulouse, France\label{aff93}
\and
Universit\'e St Joseph; Faculty of Sciences, Beirut, Lebanon\label{aff94}
\and
Departamento de F\'isica, FCFM, Universidad de Chile, Blanco Encalada 2008, Santiago, Chile\label{aff95}
\and
Department of Physics and Helsinki Institute of Physics, Gustaf H\"allstr\"omin katu 2, University of Helsinki, 00014 Helsinki, Finland\label{aff96}
\and
Departamento de F\'isica, Faculdade de Ci\^encias, Universidade de Lisboa, Edif\'icio C8, Campo Grande, PT1749-016 Lisboa, Portugal\label{aff97}
\and
Instituto de Astrof\'isica e Ci\^encias do Espa\c{c}o, Faculdade de Ci\^encias, Universidade de Lisboa, Campo Grande, 1749-016 Lisboa, Portugal\label{aff98}
\and
Cosmic Dawn Center (DAWN)\label{aff99}
\and
Niels Bohr Institute, University of Copenhagen, Jagtvej 128, 2200 Copenhagen, Denmark\label{aff100}
\and
Universidad Polit\'ecnica de Cartagena, Departamento de Electr\'onica y Tecnolog\'ia de Computadoras,  Plaza del Hospital 1, 30202 Cartagena, Spain\label{aff101}
\and
European University of Technology EUt+, European Union\label{aff102}
\and
Centre for Information Technology, University of Groningen, P.O. Box 11044, 9700 CA Groningen, The Netherlands\label{aff103}
\and
Institute of Space Sciences (ICE, CSIC), Campus UAB, Carrer de Can Magrans, s/n, 08193 Barcelona, Spain\label{aff104}
\and
Institut d'Estudis Espacials de Catalunya (IEEC),  Edifici RDIT, Campus UPC, 08860 Castelldefels, Barcelona, Spain\label{aff105}
\and
Astronomisches Rechen-Institut, Zentrum f\"ur Astronomie der Universit\"at Heidelberg, M\"onchhofstr. 12-14, 69120 Heidelberg, Germany\label{aff106}
\and
Center for Astrophysics and Cosmology, University of Nova Gorica, Nova Gorica, Slovenia\label{aff107}}    
   
\abstract{ 

We report the discovery of an ultra-faint stellar system in the vicinity of the Fornax dwarf spheroidal (dSph) galaxy in Euclid DR1. This system, which we designate as \textit{Fornax-7}, is located at \ang{1.1} from the centre of the Fornax dSph at about 145\,kpc, corresponding to a projected distance of $\sim 2.8\,\mathrm{kpc}$ assuming the distance of the Fornax dSph. We characterise Fornax-7 using constraints from its colour--magnitude diagram (CMD), \IE luminosity function (LF), and integrated optical colours from the existing ground-based data ($gri$). These observables are modelled jointly through a forward-modelling framework to constrain its distance modulus, stellar population properties, and total stellar mass. Assuming a low metallicity of $-2.2\leq\mathrm{[M/H]}\leq-2.0$, similar to the known ultra-faint systems around the Milky Way and the LMC, we derive a distance modulus consistent with that of the Fornax dSph. We estimate a stellar population age of ($10.4\pm1.9)\,\mathrm{Gyr}$, and infer a stellar mass of $170^{+50}_{-62}\,M_{\odot}$. Relaxing the metallicity constraint and allowing for a wider range that covers the metallicity of the known Fornax old globular clusters (GCs), we find a metal-poor population but with a higher metallicity $\mathrm{[M/H]} = -1.4\pm0.3$ while the age, distance modulus, and mass remain mostly the same. These results suggest a likely physical association with the Fornax dSph. Fornax-7 may represent either an extremely remote star cluster associated with Fornax, a dwarf satellite of Fornax (i.e., a “satellite of a satellite”), or an independent ultra-faint satellite in the outer halo of the Milky Way. In all cases, it constitutes an exceptional system probing the lowest-mass regime of galaxy and star cluster formation, and the structure of dwarf galaxy haloes. Deeper photometric and spectroscopic follow-up observations are required to confirm both the nature of the system and its dynamical connection to Fornax. 
}


   \titlerunning{Fornax-7, an ultra-faint companion to the Fornax dSph}
   \authorrunning{T. Saifollahi et al.}
   
   \maketitle
   
\section{\label{sc:Intro}Introduction}

The hierarchical structure formation paradigm predicts that dark matter haloes of all masses should contain populations of lower-mass subhaloes (\citealp{moore1999,springel2005,Diemand2007}). While this prediction is well established for Milky Way-mass galaxies, where numerous satellite dwarf galaxies have been observed, it should also extend to progressively lower-mass hosts (\citealp{diemand2008}). Therefore, identifying such systems is essential for testing the predictions of the cold dark matter paradigm at the lowest halo masses (\citealp{santos-santos-2022}).

Massive galaxies such as the Milky Way and M31 host numerous classical dwarf spheroidal satellites (\citealp{mateo,mcc,pace2025}). In addition, especially in recent years, numerous very low-mass stellar systems have been discovered through wide-field photometric surveys. Many of these objects, commonly classified as ultra-faint dwarf galaxies (UFDs; \citealp{willman2005,Belokurov2006,Koposov2015,martin2016,Hunter2025,joanna2026}), have stellar masses of only $10^{3}$--$10^{5}\,M_{\odot}$ and $M_V>-7.7$ (\citealp{ufd}), represent the faintest known galaxies. Their structural properties place them between Galactic globular clusters (GCs) and the classical dwarf satellite galaxies in the size--luminosity plane, blurring the traditional distinction between star clusters and galaxies (\citealp{willman2012,revaz2023,bellazini2026}). This has led to an ongoing debate regarding their true nature and whether they are dark matter-dominated dwarf galaxies or purely stellar systems. For several UFDs, medium-to-high-resolution spectroscopic observations have revealed velocity dispersions larger than expected from their stellar masses alone, suggesting the presence of dark matter (\citealp{simon2007,Strigari2008}). However, these measurements remain challenging owing to the small number of member stars, and to the possible influence of binary systems and/or other dynamical effects (\citealp{Hargreaves1996,minor2010,mc2010,gration2026,flammini2026}). These uncertainties, together with the intrinsically faint nature of UFDs at the distances of galaxies beyond the Milky Way, currently prevent robust constraints from being placed on the population of dwarf galaxy satellites around external galaxies.

The presence of satellite systems around galaxies less massive than the Milky Way and their properties remain unconstrained. Within the M31/M33 system, M33 with a stellar mass of $M_{\star,{\rm M33}} \sim 3.2 \times 10^{9}\,M_{\odot}$ (\citealp{vdm2012}), and total baryonic mass of $M_{{\rm baryon},{\rm M33}} \sim 6.4 \times 10^{9}\,M_{\odot}$, including neutral hydrogen mass (\citealp{corbelli2003}), has been suggested to host a small ultra-faint satellite system of its own (\citealp{martin2009,chipman2013,patel2018,david2022,Collins2024}). The Magellanic Clouds provide the most clear observational evidence for this hierarchical formation below the mass of the Milky Way. The Large Magellanic Cloud (LMC) with $M_{\star,{\rm LMC}} \sim 2.7 \times 10^{9}\,M_{\odot}$ (\citealp{mcc}) and $M_{{\rm baryon},{\rm LMC}} \sim 3.2 \times 10^{9}\,M_{\odot}$ (\citealp{kim1998}) hosts several candidate ultra-faint companions (\citealp{Deason2015,lmcufd1,Jethwa2016,erkal2020}), making the LMC the lowest-mass galaxy currently known to possess a confirmed ultra-faint satellite system of its own. Possible satellite systems have also been suggested around the Sagittarius dwarf spheroidal galaxy (\citealp{Luque2017}) and around several dwarf galaxies beyond the Local Group (\citealp{david2012,Crnojevic2014,annibali2016,Makarova2018,sacchi2024}), although these systems are more massive (beyond the ultra-faint regime) or have not yet been conclusively confirmed. Furthermore, while several Milky Way and Local Group dwarf galaxies show evidence of past merger events (e.g. \citealp{Amorisco2014,Kacharov2017,Cicuendez2018,Cardona-Barrero2021,Arroyo-Polonio2024}), satellites-of-satellites have not yet been directly detected at the very low-mass scales probed by ultra-faint systems. 

We have discovered a faint, compact, and highly resolved stellar system in the Euclid Data Release (DR1) $\IE$ imaging, located in the proximity of the Fornax dwarf spheroidal galaxy (Fornax dSph, \citealp{fornax-dsph}). This object, tentatively named \textit{Fornax-7}, may be physically associated with the Fornax dSph. If the association is confirmed, it could represent the first known example of an ultra-faint system, and possibly a dwarf galaxy, orbiting a host with a stellar mass of only $\sim2\times10^7\,M_{\odot}$, extending the observed hierarchical formation of satellite systems nearly two orders of magnitude below the LMC. 

The Fornax dSph is located at a heliocentric distance of about $145$\,kpc, which is corresponding to a distance modulus $(m-M)_{\rm For} = 20.8 \pm 0.1$ (\citealp{Pietrzynski2009,Oakes2022}).\footnote{Recent estimates of the distance modulus of the Fornax dSph span the range $(m-M)_{\rm For} \simeq 20.7$--$20.9$ (e.g. \citealp{Oakes2022}). Throughout this work, we adopt a distance modulus of $(m-M)_{\rm For} = 20.8 \pm 0.1$, where the uncertainty reflects the range of recent measurements reported in the literature.} Fornax has a stellar mass of $M_{*,{\rm For}} \sim 2 \times 10^7\,M_{\odot}$ ($M_{*,{\rm For}} \sim M_{{\rm baryon},{\rm For}}$) and a total mass of approximately $2 \times 10^9\,M_{\odot}$ (\citealp{mcc,errani2018,read2019}). The estimated total halo mass of the Fornax dSph depends on whether a cored or cuspy dark matter profile is assumed, spanning approximately $(0.8$--$2.5)\times10^9\,M_\odot$ (\citealp{errani2018}). 

The Fornax dSph is one of the best-studied dwarf galaxies in the Local Group owing to its unusual combination of structural and stellar population properties. The stellar body of Fornax exhibits multiple stellar populations spanning a wide range of ages and metallicities, indicating an extended and complex star formation history (\citealp{Buonanno1999,saviane2000,Battaglia2006,latarte2010,deboer2012}). The GC system of Fornax is particularly intriguing. The Fornax dSph hosts six known GCs (\citealp{hodge1961,Mackey2003,deboer2016,silvia2020,letarte2006}); five of these clusters are old ($\gtrsim10$\,Gyr) and metal poor (\citealp{strader2003,Sarajedini2024}), while Fornax-6, located near the centre of the Fornax dSph, appears to be younger, with an age of about 3\,Gyr, and more metal rich a metallicity of $-0.7$ (\citealp{chiara2026}). In addition, these GCs have an unexpectedly extended spatial distribution around the Fornax dSph despite their old ages.  

Under standard dynamical friction arguments, such massive clusters are expected to have migrated toward the centre of the galaxy over a Hubble time (\citealp{Tremaine1976}). This apparent ``timing problem'' has motivated numerous studies of the formation and dynamical evolution of the Fornax GC system and of the structure of its dark matter halo (\citealp{Oh2000,fornax4,fornax3,fornax2,fornax1}). The broad spatial distribution of Fornax GCs has frequently been interpreted as evidence for a large cored dark matter halo, disfavouring a cuspy density profile predicted by the standard cold dark matter paradigm. 

Given the relation between the number of GCs and the halo mass of their host galaxies (\citealp{harris2013}), the rich GC system would suggest to indicate that Fornax was significantly more massive in the past. In this regard, one possible scenario is that Fornax has experienced tidal interactions with the Milky Way, which may have altered both its stellar distribution and dark matter halo (\citealp{genina2022,Borukhovetskaya2022}). Alternative dynamical (e.g. modified Newtonian dynamics) and dark matter models have also been investigated to explain these observations (e.g. warm dark matter, fuzzy dark matter, \citealp{Goerdt2006,bilek2025,Szpilfidel2026}).

Given the peculiar properties of the Fornax dSph, the discovery of Fornax-7 could add another intriguing piece to this already remarkable system. Fornax-7 is located approximately \ang{1.1} from the centre of the Fornax dSph. Assuming Fornax-7 lies at the same distance as Fornax dSph, its angular separation corresponds to a projected distance of approximately $2.8$\,kpc, close to the tidal radius of the galaxy ($\sim3.27$\,kpc; \citealp{des-wang-2019}). In contrast, all previously known Fornax GCs lie within a projected distance of $\sim1.6$\,kpc of the galaxy centre. Although the current data do not allow us to determine conclusively whether Fornax-7 is a faint star cluster or a satellite dwarf galaxy of the Fornax dSph, its association to Fornax would raise important questions regarding the formation, evolution, and long-term survival of such low-mass stellar systems in that environment. Regardless of its nature, Fornax-7 may provide new insights into the assembly history, dynamical evolution, and dark matter distribution of the Fornax dSph itself. 

In this paper, we investigate the properties of Fornax-7 using the available \Euclid observations together with complementary ground-based imaging. We combine colour--magnitude diagrams (CMD), luminosity function (LF), integrated photometry, diffuse light measurements, and stochastic forward modelling to constrain its distance, age, metallicity, and stellar mass, and to assess its possible origin. The paper is organised as follows. Section~\ref{sc:data} presents the observational data and source catalogues. Section~\ref{sc:analysis} describes the analysis and stochastic forward-modelling framework. In Sect.\,\ref{sc:results} we present the results, followed by a discussion of the physical natures of Fornax-7 and its implications in the context of dwarf galaxy and star cluster formation and evolution in Sect.\,\ref{sc:dis}. We summarise the paper in Sect.\,\ref{sc:summary}. Throughout the manuscript, all magnitudes are given in the AB system.

\section{\label{sc:data} Discovery and observational data}

\begin{figure*}[h]
    \centering
    \includegraphics[width=0.49\linewidth]{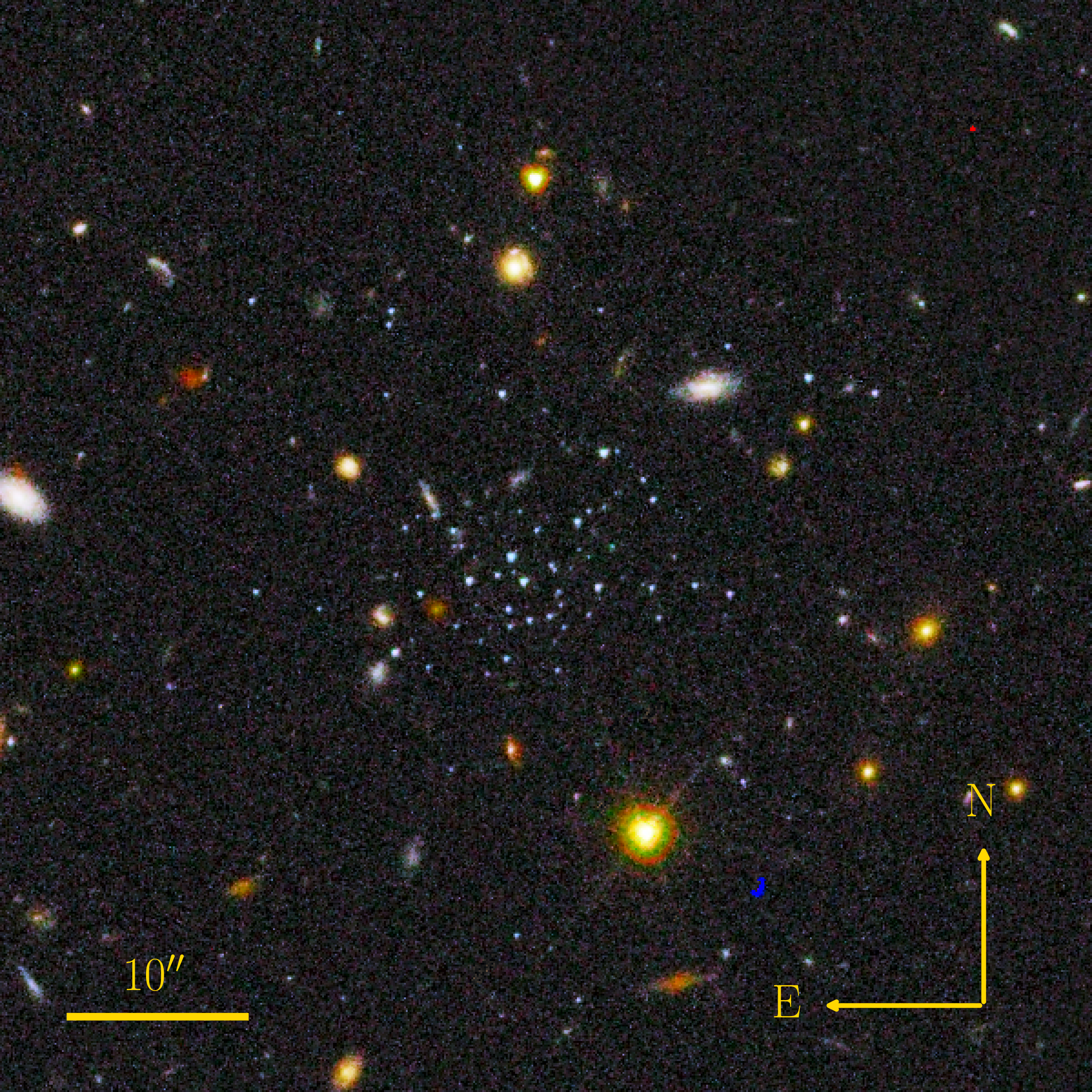}
    \includegraphics[width=0.496\linewidth]{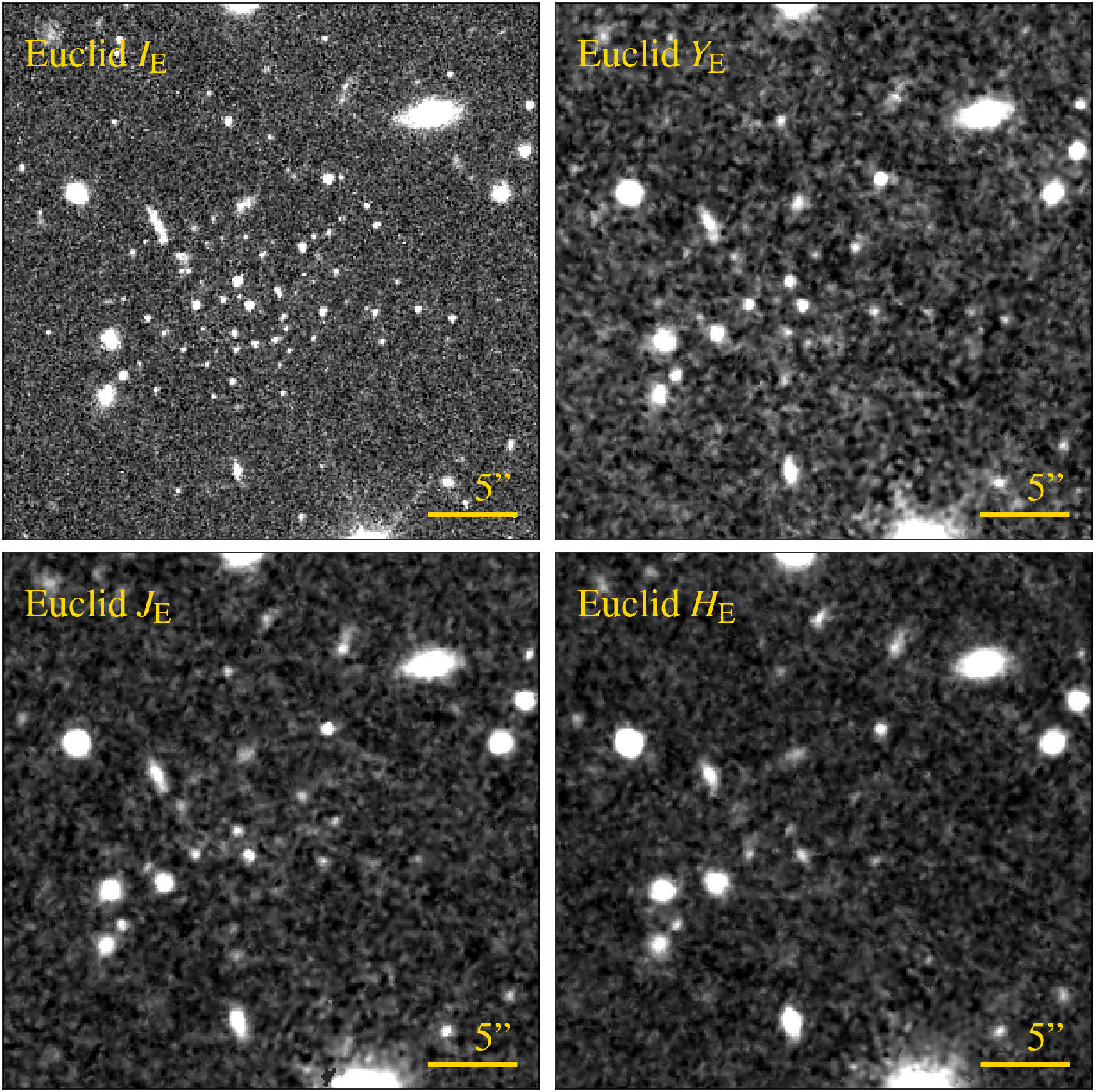}
    \caption{A \Euclid view of Fornax-7 in colour ({\it left}) and in the four \Euclid filters ({\it right}). The colour image has been created using \texttt{eummy} (\url{https://github.com/schirmermischa/eummy}), with the \IE, \YE, and \HE\ images assigned to the blue, green, and red channels, respectively.}
    \label{fig:f7-cutout}
\end{figure*}

\begin{figure}[h]
    \centering
    \includegraphics[width=\linewidth]{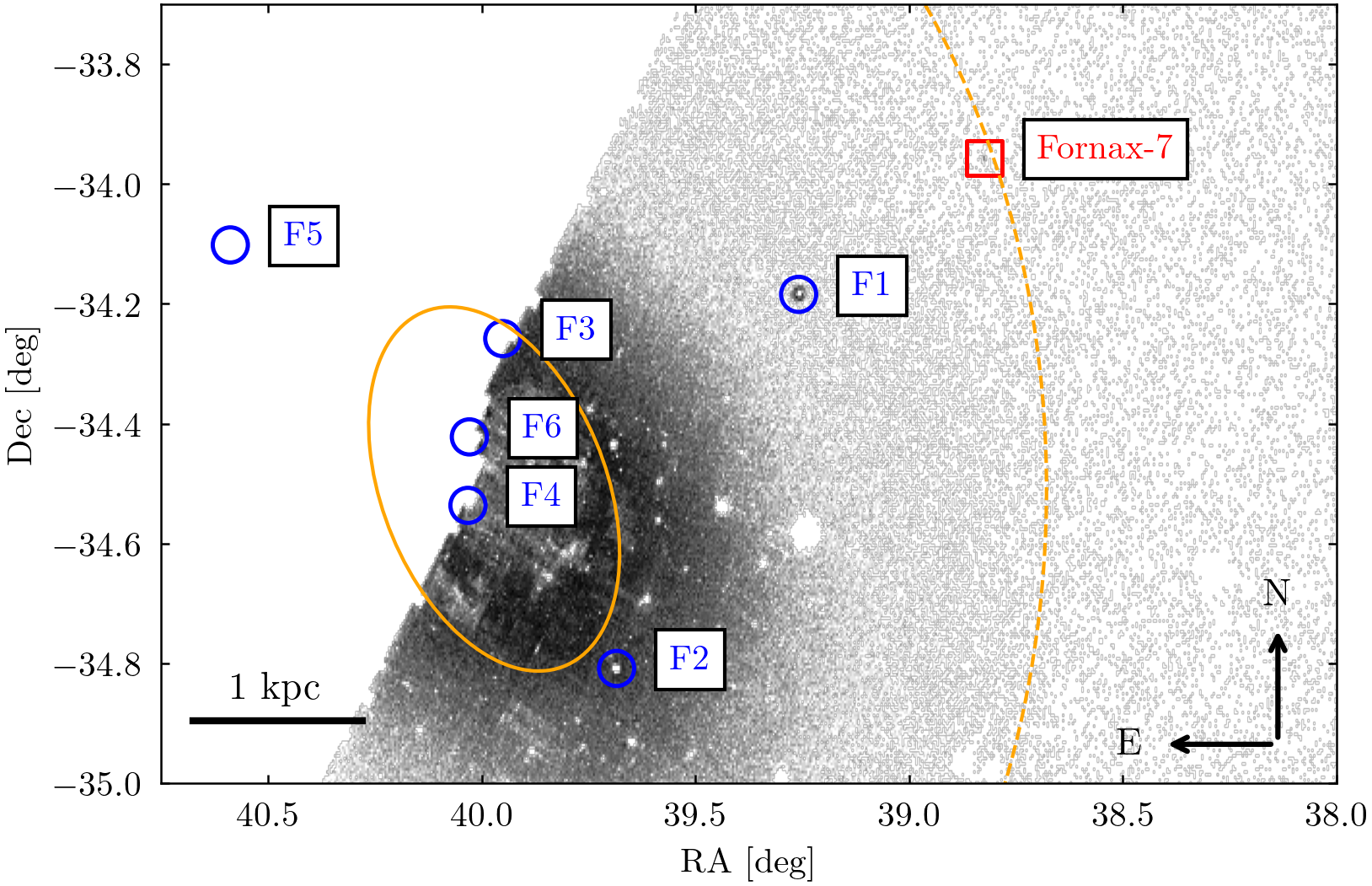}

    \caption{Location of Fornax-7 (red square) relative to the Fornax dSph and its six previously known GCs (blue circles; F1–F6). The background greyscale image shows the distribution of point-like sources detected in the Euclid DR1 catalogue (OU-MER). The solid and dashed ellipses indicate the half-light radius and tidal radius of the Fornax dSph, respectively. 
    The half-light and tidal radii of the Fornax dSph are adopted from the King-profile fit of \citet{des-wang-2019}, with $r_{\rm h}=0.88$\,kpc and $r_{\rm t}=3.27$\,kpc, respectively. The corresponding contours adopt the ellipticity 0.31, and position angle 42.2\,deg derived from the same fit.}
    \label{fig:fnx-dsph-gcs}
\end{figure}

The Fornax-7 system was identified in the internal data release of Euclid DR1 (\citealp{DR1cite}). It was serendipitously discovered as an excess of point sources, and then, through a systematic isochrone-fitting search for dwarf-galaxy candidates (\textcolor{blue}{Euclid Collaboration; Gatto et al., in prep.}). 

The Euclid DR1 (\citealp{DR1-TP001}; see also \citealp{EuclidSkyOverview} for an overview of the \Euclid mission) covers approximately 1900 deg$^2$ of the sky, primarily in the southern hemisphere. Around half of the area surrounding the Fornax dSph is covered in DR1 (see Fig.\,\ref{fig:fnx-dsph-gcs}) by the VIS instrument in the broad \IE\ band (\citealp{EuclidSkyVIS}) and by the NISP instrument in the three near-infrared \YE, \JE, and \HE\ bands (\citealp{Schirmer-EP18,EuclidSkyNISP}), together with five optical bands from ground-based imaging observations (\Euclid external data, or \texttt{EXT}). Among the \Euclid\ bands, the \IE\ observations are significantly deeper and provide substantially higher spatial resolution than the three near-infrared bands. The \IE\ imaging reaches a $5\sigma$ point-source depth of $\IE\simeq26.2$, compared to $\simeq24.3-24.5$ in the \YE, \JE, and \HE\ bands, and has a pixel scale of 0.1\arcsec and an effective angular resolution of about \ang{;;0.18}, compared to a pixel scale of \ang{;;0.3} for the NISP imaging. These characteristics were key to the identification of the Fornax-7 system, which appears as a stellar overdensity only in the \IE\ images. In the near-infrared bands, the overdensity is not readily visible beyond a handful of the brightest stars. A colour-composite image together with cutouts of Fornax-7 in all the individual \Euclid bands are shown in Fig.\,\ref{fig:f7-cutout}. The location of Fornax-7 is shown in Fig.\,\ref{fig:fnx-dsph-gcs}, together with the six known GCs of the Fornax dSph. The few brightest stars of Fornax-7 are marginally detectable in ground-based imaging from both the Kilo-Degree Survey (KiDS; \citealp{kids-dr4}) and the Dark Energy Survey (DES; \citealp{des}), however, the overdensity is not clearly visible in those data and the system has remained unidentified until the \Euclid data became available (see Fig.\,\ref{euclid-vs-ground}).

Euclid DR1 includes both images and source catalogues. In this work, we use the photometric catalogues produced from the stacked \IE image (from the OU-VIS pipeline, \citealp{DR1-TP002}), and from the final stacked tiles in all \Euclid bands (from the OU-MER pipeline, \citealp{DR1-TP009}) which are resampled to a common pixel scale of \ang{;;0.1}. The OU-VIS catalogue is based solely on the \IE observations, while the OU-MER catalogue includes source detection and photometry in all four \Euclid bands together with available ground-based ancillary data. The two catalogues are generated using different source-extraction setup configuration parameters, particularly for detection thresholds and de-blending, resulting in slight differences in the extracted source lists (see \citealp{DR1-TP002,DR1-TP009} for details). The OU-VIS source catalogues are more complete, reach fainter magnitudes, and provide improved de-blending performance, making them better suited for studying the LF and radial profile of Fornax-7. Complementary to the OU-VIS catalogue, the OU-MER catalogue provides multi-band colour information for the sources and is particularly useful for studying CMDs. The OU-VIS and OU-MER catalogues are available on the \Euclid data archive (SAS). We also incorporate optical imaging from DES in the $g$, $r$, and $i$ bands that are included in DR1 via EXT (\citealp{DR1-TP006}). These data are used to derive total magnitudes and integrated colours, providing additional constraints on the properties of the system. Throughout this work we produced several supplementary catalogues that will be made publicly available through the Centre de Données astronomiques de Strasbourg (CDS), and can also be provided upon request.

\section{\label{sc:analysis} Data analysis}

\subsection{\label{sc:star-sel}  Point-source selection}

Source catalogues were created using OU-VIS (Level 2 stacked frame catalogue) and OU-MER (Level 2 mosaic catalogue) products. The catalogues were obtained from SAS by selecting sources within 2\degree\ radius of the Fornax dSph. This area includes stars from the Fornax dSph as well as the Fornax-1 star cluster which is useful for comparisons with Fornax-7. The catalogues were combined and at this stage after being filtered for possible contamination and bad detections, as well as for extended sources, as described below. Overall, sources were selected within the magnitude range $16 < {\tt MAG\_MODEL} < 27$. 

\begin{figure}[h]
    \centering
    \includegraphics[width=\linewidth]{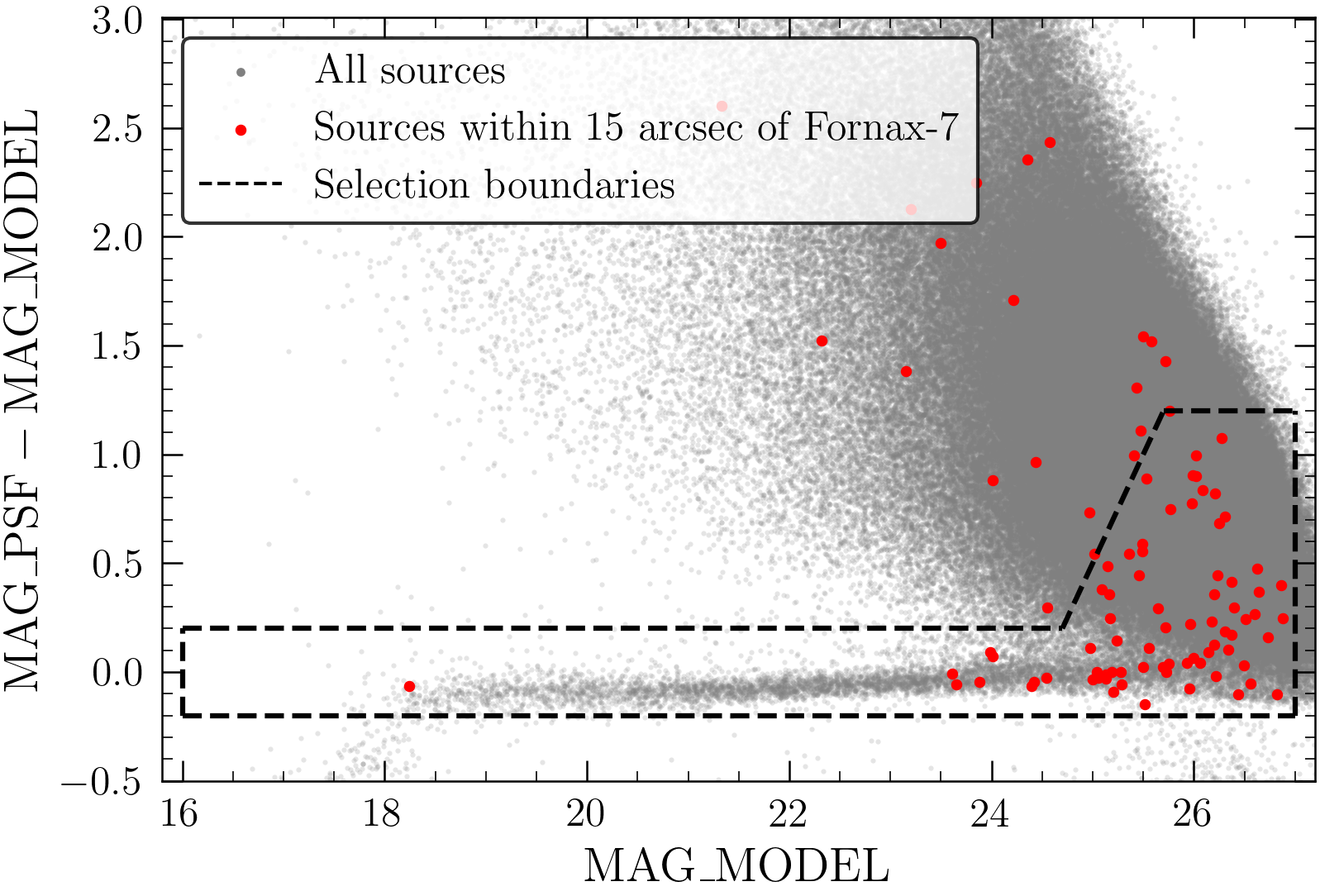}

    \caption{Selection boundaries used to identify point sources in the OU-VIS catalogue.}
    \label{selection-vis}
\end{figure}

For the OU-VIS catalogues, we use the difference between the PSF-fitting magnitude, \texttt{MAG\_PSF}, and the model-fitting  magnitude, \texttt{MAG\_MODEL}. This quantity, defined as $\Delta m = {\tt MAG\_PSF} - {\tt MAG\_MODEL}$, serves as a morphological discriminator between point-like and extended sources. While \texttt{MAG\_PSF} is derived by fitting the source with the  instrumental point spread function (PSF), \texttt{MAG\_MODEL} is obtained from an extended-source  model fit. For unresolved sources, the two measurements are expected to  be similar, whereas extended sources generally show a larger difference  between them. At bright magnitudes, compact sources were required to satisfy $-0.2 < \Delta m < 0.2$. At fainter magnitudes, where the photometric scatter increases, a progressively broader selection was adopted, requiring $-0.2 < \Delta m < 1.2$. These selection boundaries are shown in Fig.\,\ref{selection-vis}. This approach preserves faint stellar candidates while reducing contamination from extended galaxies and problematic detections. For both the OU-VIS and OU-MER catalogues, the selection criteria were calibrated to retain approximately 95\% of the point-like sources within a 21\arcsec\,radius around Fornax-7. 

For the OU-MER catalogues, only sources detected in \IE (\texttt{VIS\_DET = 1}) were considered. A cut
\texttt{POINT\_LIKE\_PROB}~>~0.01 (permissive threshold) was then applied,  in order to preferentially select compact sources consistent with stellar profiles. In addition, sources with high probabilities of being spurious detections were excluded by requiring $\texttt{SPURIOUS\_PROB} < 0.35$. The \texttt{POINT\_LIKE\_PROB} parameter gives the probability that a  detected source is point-like, with values approaching unity indicating  a high-confidence unresolved source, while \texttt{SPURIOUS\_PROB}  quantifies the probability that a detection is spurious rather than a  genuine astrophysical source. 

\subsection{\label{sc:light-profile}  Structural properties}

The radial profile of Fornax-7 was constructed using the selected OU-VIS point-like sources within the magnitude range $16.0 < \IE < 26.5$. Starting from an initial estimate of the system centre, we constrained the structural parameters of Fornax-7 using a maximum-likelihood analysis of the spatial distribution of stars with a constant foreground/background (\citealp{martin2008}). We obtain a half-light radius\footnote{This parameter formally represents the half-count radius of the system. As noted by \citet[][footnote~4]{martin2008}, in the absence of significant mass segregation, the half-count radius can also be interpreted as an estimate of the half-light radius.} of $R_{\rm h} = \mbox{\ang{;;7.12}} \pm \mbox{\ang{;;1.49}}$. The inferred position angle is ${\rm PA}=116^{+22}_{-30}$\,deg, while the ellipticity is $\epsilon=0.02^{+0.35}_{-0.02}$, consistent with a nearly circular morphology but with a poorly constrained upper limit. The coordinates of the centre are $\alpha$ = \ra{02;35;17.94} and $\delta$ = \ang{-33;57;24.25}. Assuming Fornax-7 is at the same distance as the Fornax dSph, we measure a half-light radius of $R_{\rm h}=(5.0\pm1.0)$\,pc.

\subsection{\label{sc:lf}  Luminosity function (LF)}

\begin{figure}
    \centering
    \includegraphics[width=0.9\linewidth]{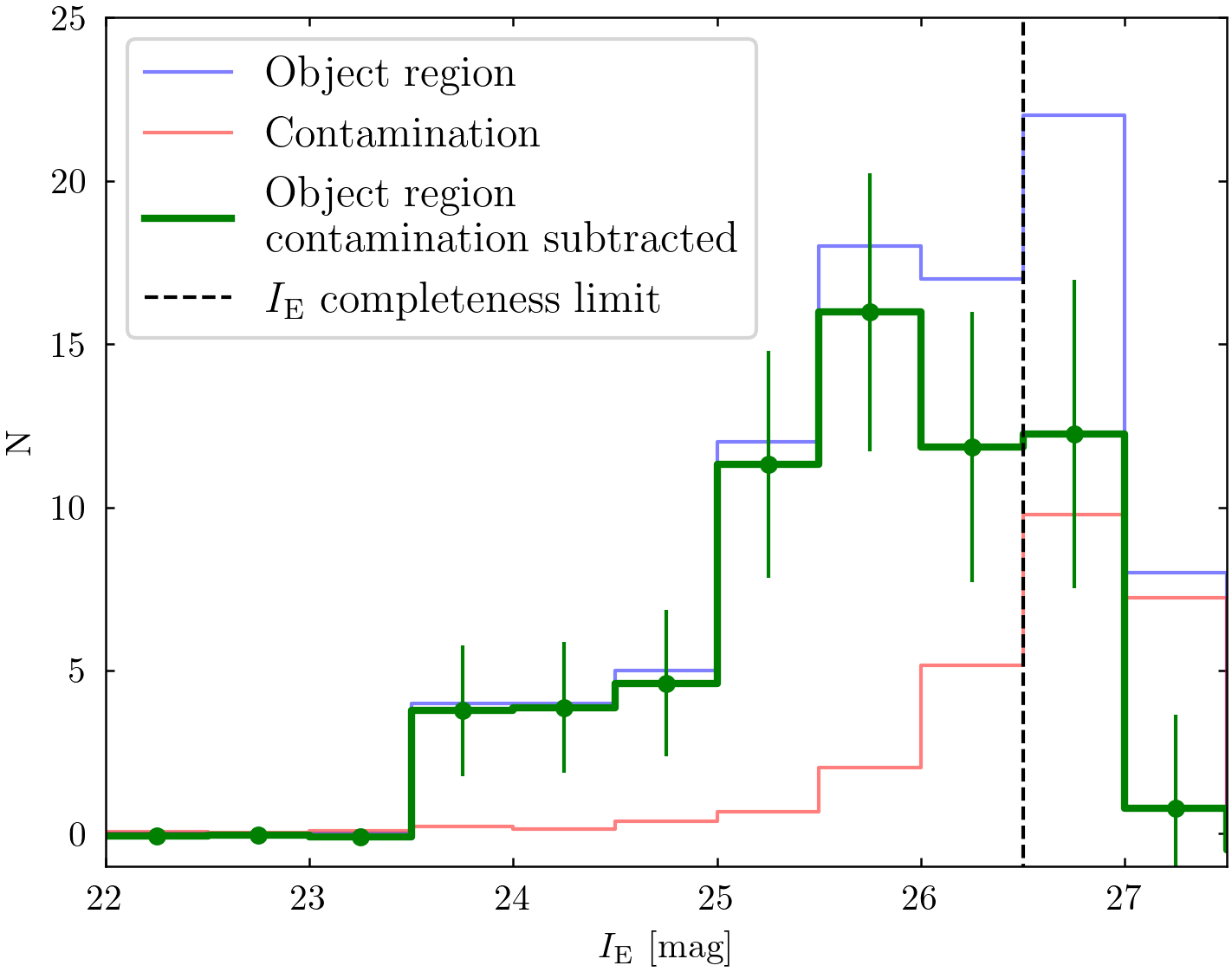}
    \caption{LF of Fornax-7 corrected for foreground/background contamination (green). The blue and red histograms show point-like sources within 21\arcsec of the Fornax-7 centre and the estimated contamination, respectively. The black dashed line marks the expected 80\% completeness limit of the catalogue in the \IE band. Error bars include Poisson uncertainties in both source counts and estimated contamination.}
    \label{fig:f7-lf}
\end{figure}

The LF of Fornax-7 (Fig.\,\ref{fig:f7-lf}) was constructed using the selected OU-VIS point-like sources within a circular region of radius 21\arcsec\,centred on the system, which corresponds to approximately three times the estimated $R_{\rm h}$. To account for foreground and background contamination, a local foreground/background LF (hereafter contamination LF) was estimated from an annular region between 30\arcsec\,and 120\arcsec\,from the centre. The contamination counts were scaled according to the relative areas of the object and contamination regions, and subsequently subtracted from the LF measured within the central aperture. The LF was constructed in the $\IE$ band using bins of width 0.5. The total integrated luminosity of Fornax-7 was estimated from the contamination-subtracted LF by converting the net counts in each magnitude bin into flux units and summing the contributions over the magnitude range $22 < \IE < 26.5$, corresponding to the regime where the LF remains reliable and sufficiently complete, more than 80\% (according to the completeness analysis in \citealp{saifollahi2025a} and \citealp{urbano2025}).

The total $\IE$ magnitude of Fornax-7 was estimated by combining the flux from the resolved stellar population with that of the unresolved diffuse component. For the resolved stars brighter than $\IE=26.5$, we obtain an integrated magnitude $\IE=20.95$. To account for the contribution from stars fainter than this limit, all detected sources with $\IE<26.5$~mag, which were already included in the resolved stellar luminosity estimate, were masked in the $\IE$ image using circular masks scaled according to their measured source sizes. Then, the local sky flux was estimated from regions located beyond $3\,R_{\rm h}$ and subtracted from the image. The residual flux was then measured within one half-light radius of the system, excluding the masked regions. Assuming that half of the total light from those fainter stars is enclosed within $R_{\rm h}$, the measured flux was multiplied by a factor of two, yielding an integrated magnitude of $\IE=20.8 \pm 0.2$ for the total \IE magnitude of Fornax-7. Note that this estimate assumes that the unresolved faint stars follow the same spatial distribution as the brighter resolved stars and does not account for possible mass segregation. We also estimated the surface brightness of the diffuse light. After masking all detected sources, we measured a mean surface brightness of $\mu_{\rm e} \sim 27.7$ mag\,arcsec$^{-2}$ within the half-light radius, indicating the presence of an extremely faint stellar component below the direct source-detection limit. 

\begin{figure*}[h!]
    \centering
    \includegraphics[width=0.9\linewidth]{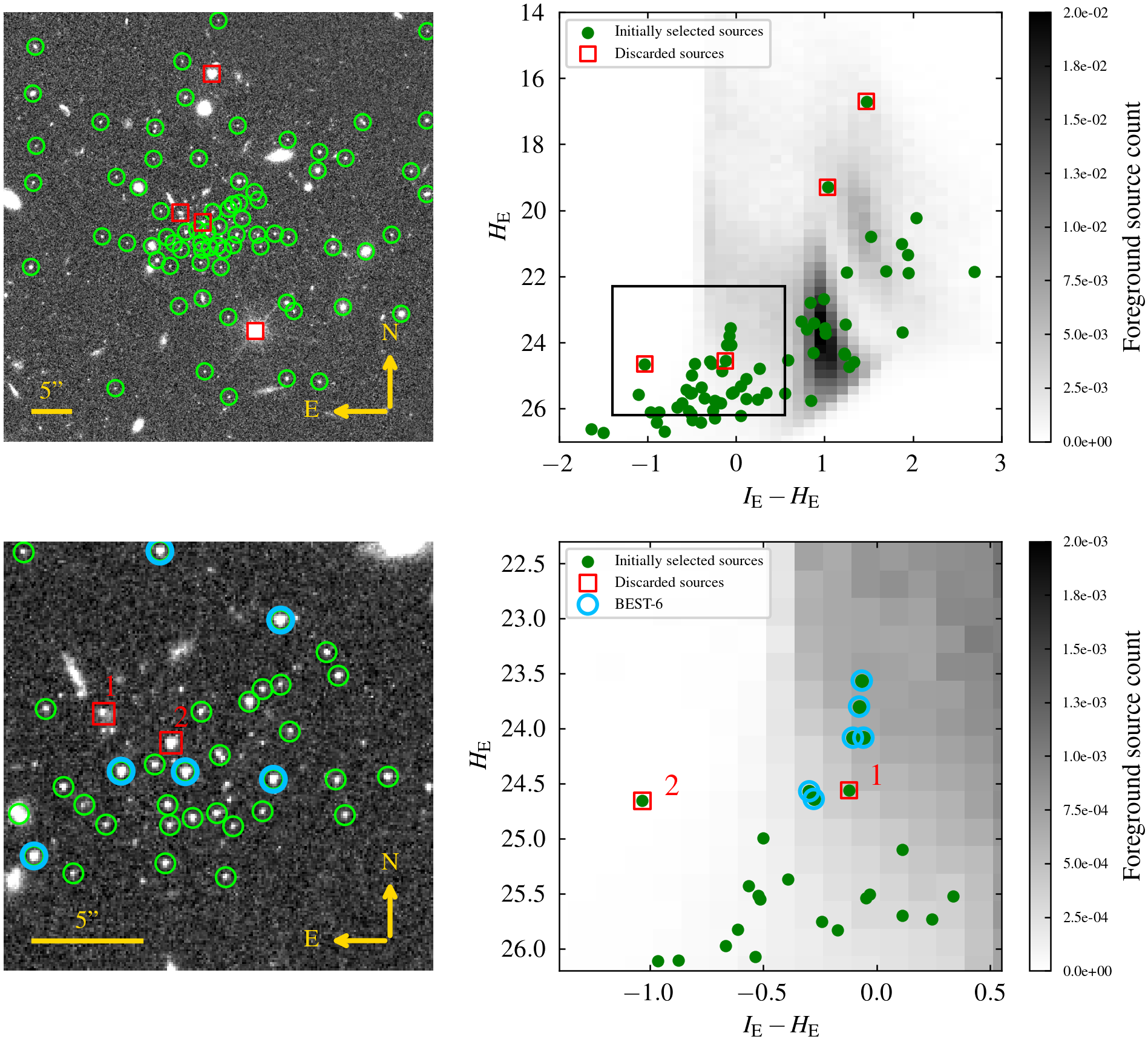}
    \caption{Spatial distribution of point-sources (in OU-MER catalogue) within \IE cutouts of 60” ({\it top-left}) and 30” ({\it bottom-left}). The {\it right} panels show the corresponding CMDs ($\IE-\HE$ versus $\HE$). Green circles indicate the initially selected point-sources, while red squares mark those that were excluded. The grayscale background indicates the expected foreground contamination from the Besançon Galactic models, convolved with the photometric errors (no completeness limit is applied). Sources with $\IE-\HE > 0.5$ are consistent with foreground Milky Way stars and are also excluded from the analysis. The final candidates used for CMD analysis (BEST-6) are highlighted with cyan circles. The black box in the \textit{top-right} panel indicates the region shown in the \textit{bottom-right} panel.}
    \label{fig:f7-det-2}
\end{figure*}

\begin{figure}[h!]
    \centering
    \includegraphics[width=\linewidth]{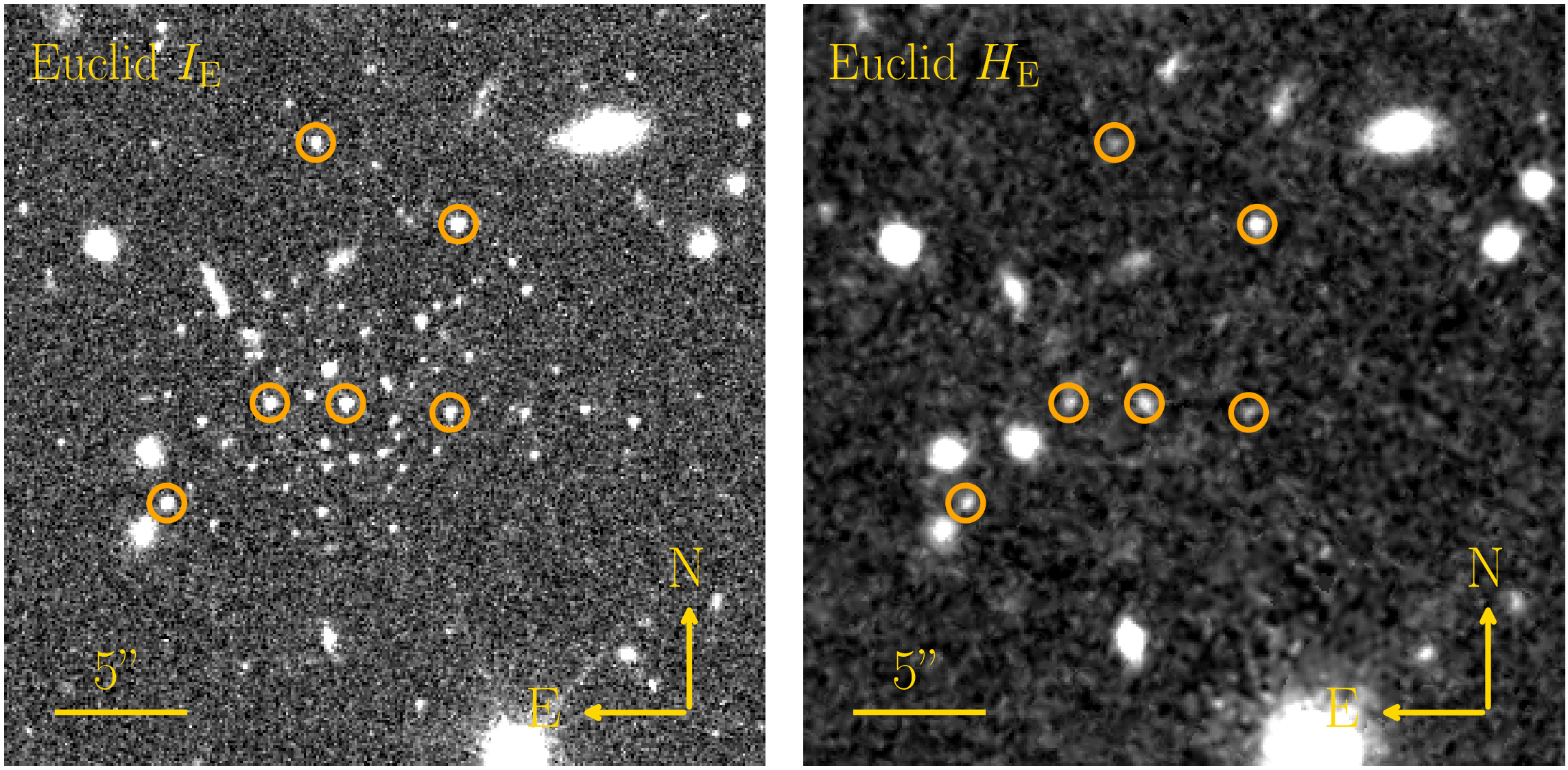}
    \caption{\IE\ ({\it left}) and NISP-\HE\ ({\it right}) cutouts of the central region of Fornax-7. The orange circles indicate the six candidate member stars (BEST-6) selected for the CMD analysis. These objects represent the brightest sources with reliable $\IE-\HE$ colours after excluding foreground probable contaminants and problematic detections.
    }
    \label{fig:f7-best-6}
\end{figure}

\subsection{\label{sc:cmd}  Colour--magnitude diagram (CMD)}

\begin{figure*}
    \centering
    \includegraphics[width=0.24\linewidth]{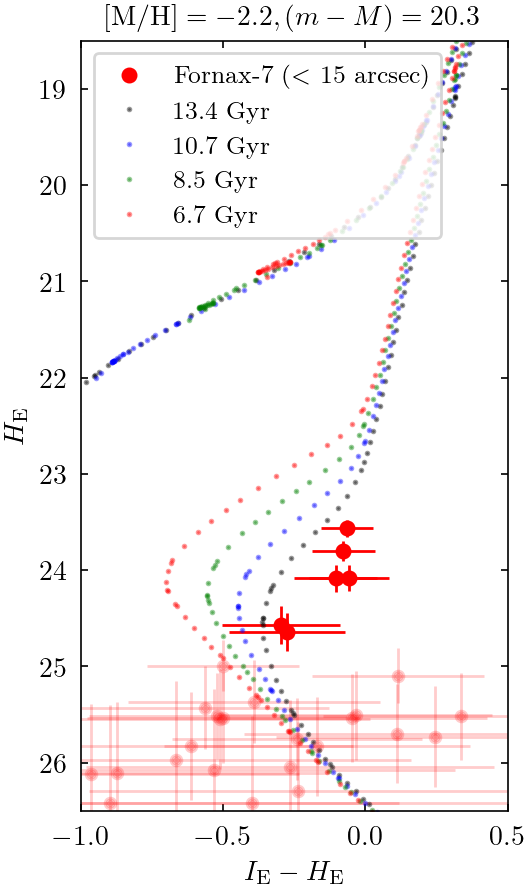}
    \includegraphics[width=0.24\linewidth]{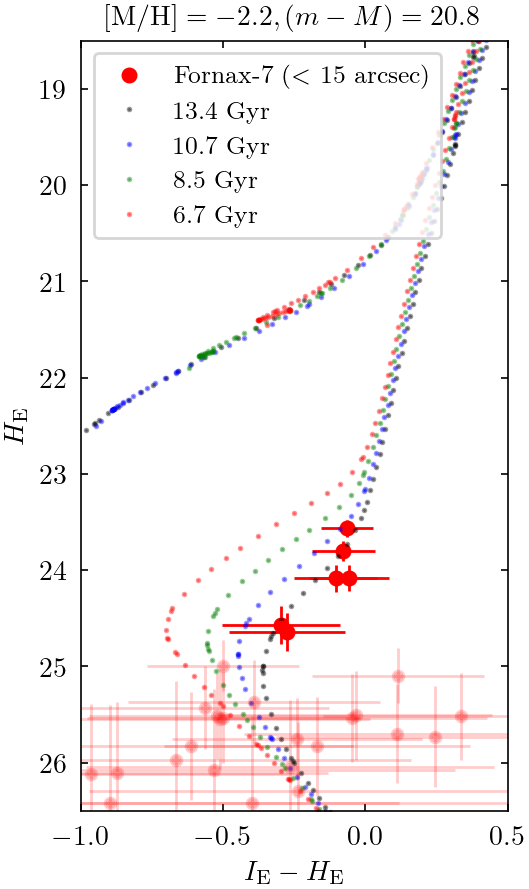}
    \includegraphics[width=0.24\linewidth]{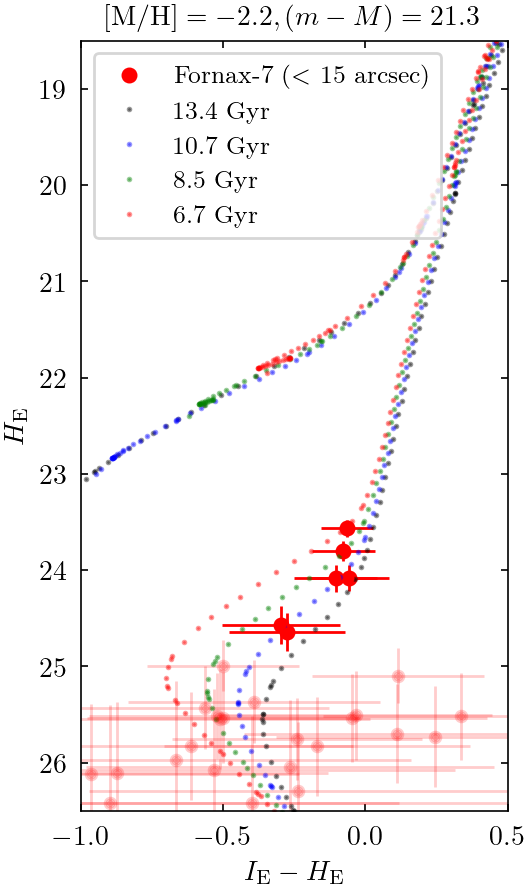}
    \includegraphics[width=0.24\linewidth]{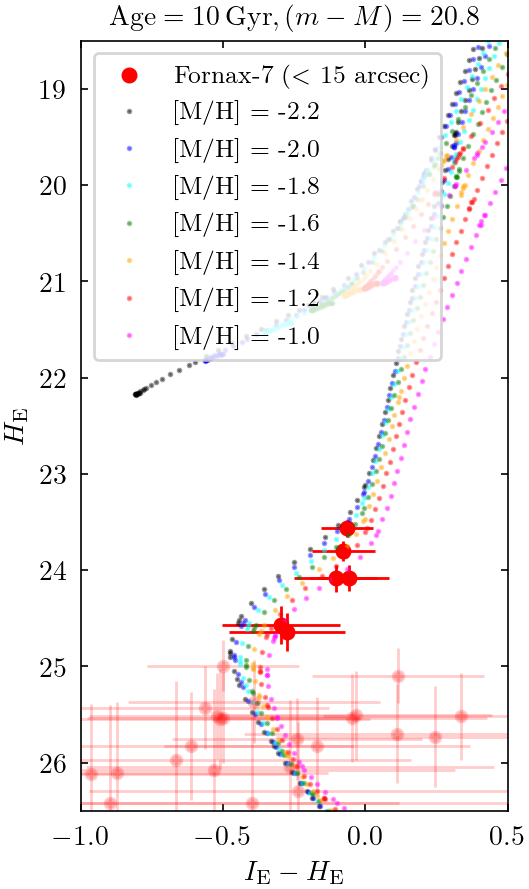}
    \caption{CMD of Fornax-7 stars (red, with errorbars) including the BEST-6 (larger symbols) compared with PARSEC isochrones. In the first three panels (from left to right) the isochrones refer to a constant metallicity, different ages and three distance moduli, respectively, while in the fourth panel show a constant age and different metallicities.}
    \label{fig:f7-cmd}
\end{figure*}

\begin{figure}
    \centering
    \includegraphics[width=0.9\linewidth]{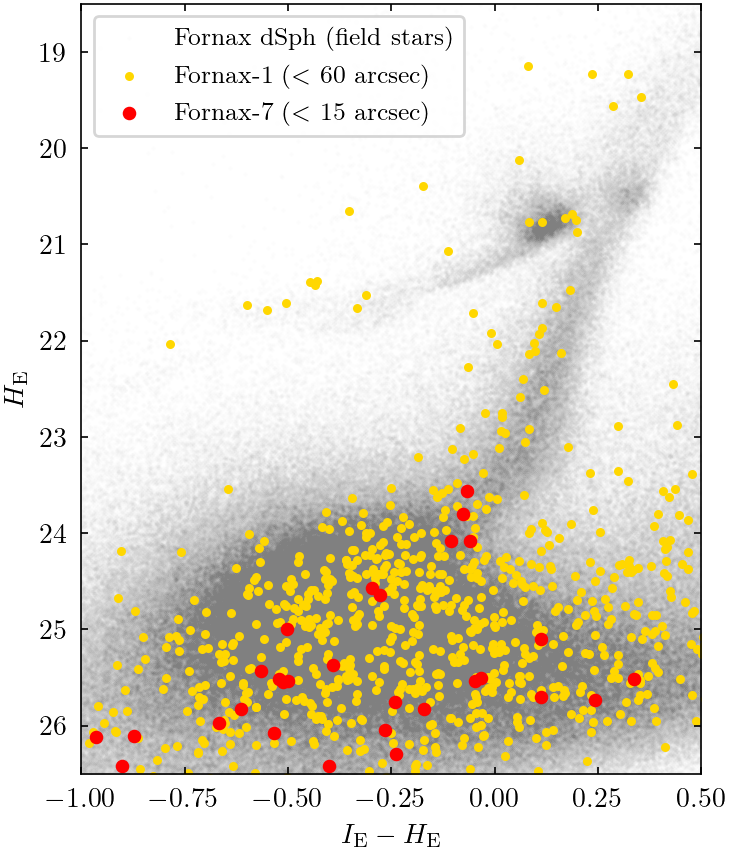}
    \caption{CMD of Fornax-7 stars (red) with stars from the Fornax dSph field population (grey) and the GC Fornax-1 (yellow).}
    \label{fig:f7-cmd-f1}
\end{figure}

Here we use the OU-MER catalogues to construct the CMD of Fornax-7. Owing to the limiting depth of the \HE\ observations, reliable colours are only available for sources brighter than $\HE \sim25$, while fainter sources are dominated by photometric scatter. Figure \ref{fig:f7-det-2} shows the detected sources together with their CMDs in $\IE-\HE$ versus $\HE$ for cutouts of 60\arcsec\,and 30\arcsec. The figure also includes the expected contamination from foreground Milky Way stars based on the Besançon Galactic models (\citealp{robin2004,Czeka2014,robin2022}). Sources with $\IE-\HE > 0.5$ are consistent with foreground stellar contamination and are therefore excluded from the subsequent analysis. 

A number of additional sources, indicated by red squares in Fig.\,\ref{fig:f7-det-2} were manually removed. The two brightest objects have \textit{Gaia} proper motions consistent with Milky Way stars and would otherwise be incorrectly interpreted as possible RGB stars associated with Fornax-7. These two stars are the only stars within the 60\arcsec\ box region centred on Fornax-7 with astrometric measurements available from \textit{Gaia}. Their  parallaxes are $(0.863\pm0.982$)~mas and  $(2.719\pm0.147)$~mas, respectively. The first star has proper  motions of $\mu_{\alpha}=(10.983\pm0.552)$~mas~yr$^{-1}$ and  $\mu_{\delta}=(-4.153\pm0.972)$~mas~yr$^{-1}$, corresponding to  significances of $\sim20\sigma$ and $\sim4\sigma$, respectively, while the second has $\mu_{\alpha}=(57.412\pm0.085)$~mas~yr$^{-1}$  and $\mu_{\delta}=(5.074\pm0.139)$~mas~yr$^{-1}$, corresponding to  $\sim675\sigma$ and $\sim36\sigma$. Their significant proper motions indicate that both objects are foreground Milky Way stars rather than members of Fornax-7.

Two additional sources with $\IE-\HE < 0.5$ were also excluded. One object with extremely blue colour is likely either a foreground white dwarf or a compact background galaxy. The other object appears to be blended with an extended background source in the \IE\ images, preventing a reliable photometric measurement. After applying these selections, six stars brighter than $\HE \sim 25$ with reasonably constrained colours (magnitude errors less than 0.25) remain (shown as blue symbols in the lower panel). These six candidate member stars are also displayed in the \IE\ and \HE\ cutouts in Fig.\,\ref{fig:f7-best-6}. Hereafter, we refer to these six stars as the \textit{BEST-6}.

The candidate Fornax-7 stars are compared in Fig.\,\ref{fig:f7-cmd} with PARSEC isochrones (\citealp{parsec})  assuming $[{\rm M/H}] = -2.2$ at distance moduli of 20.8 (consistent with the Fornax dSph), 20.3 (as example for a closer object), and 21.3 (as example for a more distant object), with an extinction of $A_V=0.05$~mag.\footnote{\url{https://irsa.ipac.caltech.edu/applications/DUST/}} At a distance modulus of 20.8, older isochrones provide a better match to BEST-6, within their photometric uncertainties, whereas at larger distances younger stellar populations become more consistent with the observed CMD. This illustrates the strong age--distance degeneracy present in the data for a metal-poor stellar population. At old ages, variations in metallicity of the order $\sim0.2$\,dex produce only modest changes in the isochrone positions in this colour space (see the {\it right} panel in Fig.\,\ref{fig:f7-cmd}). 

Figure\,\ref{fig:f7-cmd-f1} compares the CMD of Fornax-7 with stars from the Fornax dSph (grey points) and the GC Fornax-1 (yellow points). The stellar population of the Fornax dSph spans a broad range of ages and metallicities, producing a wider and generally redder CMD distribution (\citealp{Buonanno1999,saviane2000,deboer2012}). In contrast, Fornax-1 is old and metal poor (\citealp{strader2003,Sarajedini2024}), and its CMD morphology appears more similar to that of Fornax-7. Inspecting these diagrams visually, the Fornax-7 BEST-6 are seemingly consistent with subgiant stars near the main-sequence turn-off for Fornax-1 as well as the Fornax dSph, while no convincing Red Giant Branch (RGB) population is detected. Additional CMDs in other colour combinations, including $\IE-\YE$ and $\IE-\JE$, are presented in Fig.\,\ref{fig:f7-cmd-extra}. Although the Fornax-7 CMD and the BEST-6 provide initial constraints on the stellar population of Fornax-7, the CMD alone is not sufficiently constraining because of the small number of detected sources and the absence of bright evolved stars. In fact, that lack of bright stars is itself an interesting constraint on the population. Additional constraints from the LF and integrated colours are incorporated through the forward-modelling analysis presented in the next subsection.

\subsection{\label{sec:int_colours} Integrated colours}

The integrated photometry of Fornax-7 was estimated using the ground-based \Euclid-EXT imaging in the $g$, $r$, and $i$ bands. Owing to the lower spatial resolution of these data compared to the \Euclid \IE images, extended background sources identified in the higher-resolution \IE imaging, as well as the rejected point sources (indicated by red squares in Fig.\,\ref{fig:f7-det-2}) were masked prior to the photometric measurements. Aperture photometry was then performed in each band by constructing a curve of growth, and the total integrated magnitude was taken from the asymptotic value where the enclosed flux reached a plateau. From these measurements, we derive integrated colours of $g-r = 0.25\pm0.1$ and $g-i = 0.4\pm0.1$. The uncertainties reflect the scatter among the measurements
in the plateau region, as well as the robustness of the photometry by varying the background region and source-masking procedure, resulting in colour variations of order 0.1. As an independent check on the integrated colours, we also considered the median colours of the BEST-6. For stars in the subgiant branch and at lower masses, the corresponding isochrone colours vary within about 0.2 over the relevant range, such that the median colours of the BEST-6 would provide a useful estimate of the underlying stellar population. For the BEST-6, we obtain $g-r \simeq 0.38$ and $g-i \simeq 0.49$~mag. Given these measurements, for the analysis of this paper, we adopt representative colours of $g-r = 0.3\pm0.1$ and $g-i = 0.45\pm0.1$.

\subsection{\label{sec:forward_modeling} Forward-modelling}

Our forward-modelling is based on stochastic realisations of synthetic stellar populations. The explored parameter space is defined by $M = (\tau,\,[{\rm M/H}],\,\mu)$, where $\tau$ is the stellar age, $[{\rm M/H}]$ is the metallicity, and $\mu$ is the distance modulus. First, we restrict the metallicity to $-2.2 \leq \mathrm{[M/H]} \leq -2.0$, consistent with the metal-poor populations of the low-mass dwarf galaxies with $M_\star < 10^5 M_{\odot}$ (\citealp{kirby2013}), and most known ultra-faint systems (\citealp{willman2012,cerny}). The explored parameter grid covers old and metal-poor populations with\\
\\
$9.83 \leq \log_{10}({\rm{\it \tau}/yr}) \leq 10.13$\,,\\
$-2.2 \leq [{\rm M/H}] \leq -2.0$\,,\\
$19.8 \leq \mu \leq 21.8$\,.\\
\\
The main results presented in this paper are based on this strict metallicity constraint. To assess the dependence of the inferred age and distance modulus on this assumption, we later relax the metallicity constraint by allowing\\
\\
$-2.2 \leq [{\rm M/H}] \leq -1.0$\,.\\
\\
This range of metallicities encompasses
the metallicities of the known Fornax dSph GCs (\citealp{strader2003,Sarajedini2024}), excluding the relatively metal-rich Fornax-6 with [M/H]=$-0.7$ (\citealp{pace2021,chiara2026}) that reside near the centre of the Fornax dSph. Furthermore, we investigated both Kroupa and Chabrier initial mass function (IMF; \citealp{kroupa,chabrier}). 

In this work, PARSEC isochrones were adopted. The isochrones were downloaded from the PARSEC web interface\footnote{\url{https://stev.oapd.inaf.it/cgi-bin/cmd}}. We adopted the default parameters, with an extinction of $A_V=0.05$~mag. Isochrones were generated over a grid of ages sampled uniformly in 0.025\,dex. For the metallicity, we adopted a sampling of 0.05\,dex for the strict metallicities, and 0.10\,dex for the relaxed metallicities.

\begin{figure}
    \centering
\includegraphics[width=0.9\linewidth]{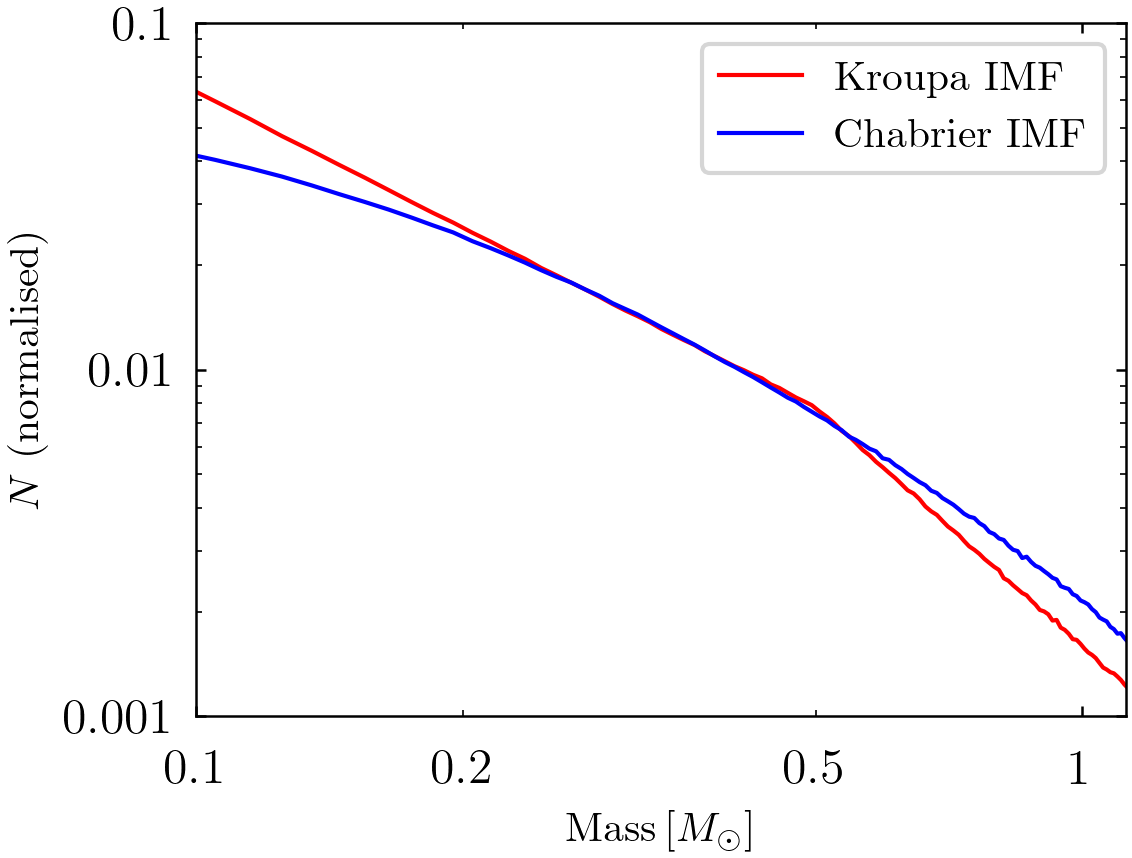}
    \caption{Comparison of the normalised stellar mass distributions for the Kroupa (red) and Chabrier (blue) IMFs.}
    \label{fig:IMFs}
\end{figure}

For each combination of age, metallicity, and distance modulus, stellar populations were generated through random sampling of the IMF. Individual initial stellar masses were drawn randomly from either the Kroupa or Chabrier IMF. Stars were sampled from the adopted IMF with a lower-mass limit of
$0.1\,M_\odot$. However,
for the range of old stellar populations considered, the adopted isochrones naturally contain surviving stars with initial masses only up to approximately $1.1\,M_\odot$ (due to the old age of the stellar population). Figure~\ref{fig:IMFs} compares the Kroupa and Chabrier IMFs over the mass range $0.1$--$1.1\,M_\odot$. The two distributions are broadly similar, although the Kroupa IMF leads to a somewhat larger fraction of low-mass stars,
whereas the Chabrier IMF contains relatively more stars toward the high-mass end of this interval.

For each simulated star, an initial stellar mass was randomly drawn from the adopted IMF. The corresponding magnitudes at the specified age and metallicity were then obtained by linearly interpolating the PARSEC isochrone as a function of initial stellar mass. Magnitudes were generated in the \Euclid\ bands \IE, \YE, \JE, and \HE bands, as well as the EXT bands $g$, $r$, and $i$. For each realisation, we applied a number of constraints, and generate a number of observables which were then compared with the measurements presented in the previous sections. These are:

\begin{enumerate}
\item the absence of bright stars (e.g. RGB), brighter than the adopted threshold magnitude ($\IE \leq 23$),
\item the \IE-band total magnitude of stars in the range $23 \leq \IE \leq 26.5$ within 0.5 of the one of Fornax-7,
\item the total \IE luminosity of the faint/diffuse unresolved stars ($\IE > 26.5$) within 50\% of the observed value,
\item the $\IE-\HE$ versus $\IE$ CMD,
\item the \IE-band LF,
\item the integrated optical colours $g-r$ and $g-i$.
\end{enumerate}

Constraints~1--3 are are treated as hard constraints, through a probability, which we refer to as the occurrence probability, $P_{\rm occ}$. We evaluate $P_{\rm occ}$ as the fraction of our random realisations that satisfy them. To what degree the remaining three constraints are satisfied is evaluated through three $\chi^{2}$ statistics, corresponding to the CMD, LF, and integrated colours (CL). These are then combined with $P_{\rm occ}$ into the metric $Q$, defined as
\begin{equation}
Q = \chi^2_{\rm CMD} + \chi^2_{\rm LF} + \chi^2_{\rm CL} -2\ln(P_{\rm occ})\,,
\label{eq1}
\end{equation}
where lower values of $Q$ correspond to models that simultaneously satisfy the hard constraints and provide a better match to the observed photometric properties. We identify the best-fitting models by minimizing this quantity (see Appendix\,\ref{app:baysian} for the details of the Bayesian description of this metric).

For each combination of age, metallicity, and distance modulus, we generate realisations containing between 0 and 2000 stars, sampled from the adopted IMF down to a minimum stellar mass of $0.1\,M_\odot$. We adopted the upper limit of 2000 stars given the expected stellar mass of Fornax-7. For a present-day stellar mass of $\sim250\,M_\odot$ and an average stellar mass of $\sim0.25\,M_\odot$, the system is expected to contain $\sim1000$ stars. Therefore, an upper limit of 2000 stars covers the expected range and more, while avoiding unnecessary computational expense.

The sampling procedure is repeated until either 1000 successful realisations, satisfying constraints~1--3, have been obtained or a maximum of 5000 unique realisations has been generated. The latter limit is imposed because, in some regions of parameter space, particularly those predicting luminous evolved stars, fewer than 1000 (and in some cases none) of the realisations satisfy constraints~1--3 before reaching this limit. Then, we estimate the occurrence probability, $P_{\rm occ}$, as the ratio of the number of successful realisations satisfying constraints 1--3 to the total number of stochastic realisations generated. Thus, $P_{\rm occ}$ measures the probability that a stellar population reproduces the primary observed properties of Fornax-7. 

For each combination of age, metallicity, and distance modulus, the corresponding PARSEC isochrone was shifted to the trial distance modulus. The isochrone was then interpolated to predict the expected colour at the observed magnitudes of the BEST-6 stars in \HE. Then, the CMD $\chi^2$ statistic was computed by comparing the observed and model colours, using
\begin{equation}
\chi^2_{\rm CMD}
=
\sum_{i=1}^{6}
\frac{\left(C_{{\rm obs},i}-C_{{\rm model},i}\right)^2}
{\sigma_{C,i}^2}\,,
\end{equation}
where $C=\IE-\HE$ is the colour, $C_{{\rm obs},i}$ and $\sigma_{C,i}$ are the observed colour and its uncertainty for the $i$th star, and $C_{{\rm model},i}$ is the interpolated colour predicted by the isochrone at the same $\HE$ magnitude. Then, the LF $\chi^2$ statistic, $\chi^2_{\rm LF}$, was quantified using
\begin{equation}
\chi^2_{\rm LF} =
\sum_i
\frac{
\left(
N_{{\rm model},i} - N_{{\rm obs},i}
\right)^2
}
{\sigma_{{\rm LF},i}^2}\,,
\end{equation}
where $N_{{\rm model},i}$ and $N_{{\rm obs},i}$ are the model and observed numbers of stars in each magnitude bin, and $\sigma_{{\rm LF},i}$ is the uncertainty of the observed \IE LF. Finally, the integrated-colour $\chi^2$ statistic, $\chi^2_{\rm CL}$, was computed by comparing the observed integrated $g-i$ and $g-r$ colours, derived from the ground-based imaging, with the corresponding colours predicted by each model
\begin{equation}
\chi^2_{\rm CL} =
\frac{
\left[(g-r)_{\rm obs} - (g-r)_{\rm model}\right]^2
}{
\sigma_{g-r}^2
}
+
\frac{
\left[(g-i)_{\rm obs} - (g-i)_{\rm model}\right]^2
}{
\sigma_{g-i}^2
}\,,
\end{equation}
where $(g-r)_{\rm obs}$ and $(g-i)_{\rm obs}$ are the observed integrated EXT colours of Fornax-7, $(g-r)_{\rm model}$ and $(g-i)_{\rm model}$ are the corresponding colours predicted by the stochastic stellar-population models, and $\sigma_{g-r}$ and $\sigma_{g-i}$ are the observational uncertainties on the integrated colours. For the forward modelling described here, the $\chi^2$ statistics ($\chi^2_{\rm CMD}$, $\chi^2_{\rm LF}$, and $\chi^2_{\rm CL}$) are computed only for realisations satisfying constraints 1--3. Then, the metric $Q$ was computed as described. To derive the final parameter estimates, the metric $Q$ was converted into relative likelihood weights according to
\begin{equation}
w_i \propto
\exp\left[
-\frac{1}{2}(Q_i - Q_{\rm min})
\right]\,,
\end{equation}
where $Q_{\rm min}$ is the minimum value of the metric. The weights were then used to compute the weighted mean (first moment) and the weighted covariance matrix (second central moment) of the parameters. The weighted means were adopted as the best-fitting parameter values, while the covariance matrix provides the corresponding uncertainties and parameter correlations.

The likelihood construction in this work assumes that the CMD, LF, and the integrated $gri$ colour constraints can be treated as independent. In practice, these observables are derived from the same underlying stochastic stellar population and may therefore exhibit some covariance. A full propagation of these dependencies would require dedicated numerical experiments that are beyond the scope of the present work. We have, however, tested several variations of the inference procedure, including changes in the relative weighting of the different terms and in the adopted selection criteria, and find that the resulting changes in the inferred parameters consistently remain within the quoted uncertainties. We therefore regard the inferred parameter uncertainties as a reasonable representation of the robustness of the present analysis, while noting that they should not be interpreted as accounting for all possible systematic correlations between the observables.

\section{\label{sc:results} Results}

\begin{figure}
    \centering
    \includegraphics[width=\linewidth]{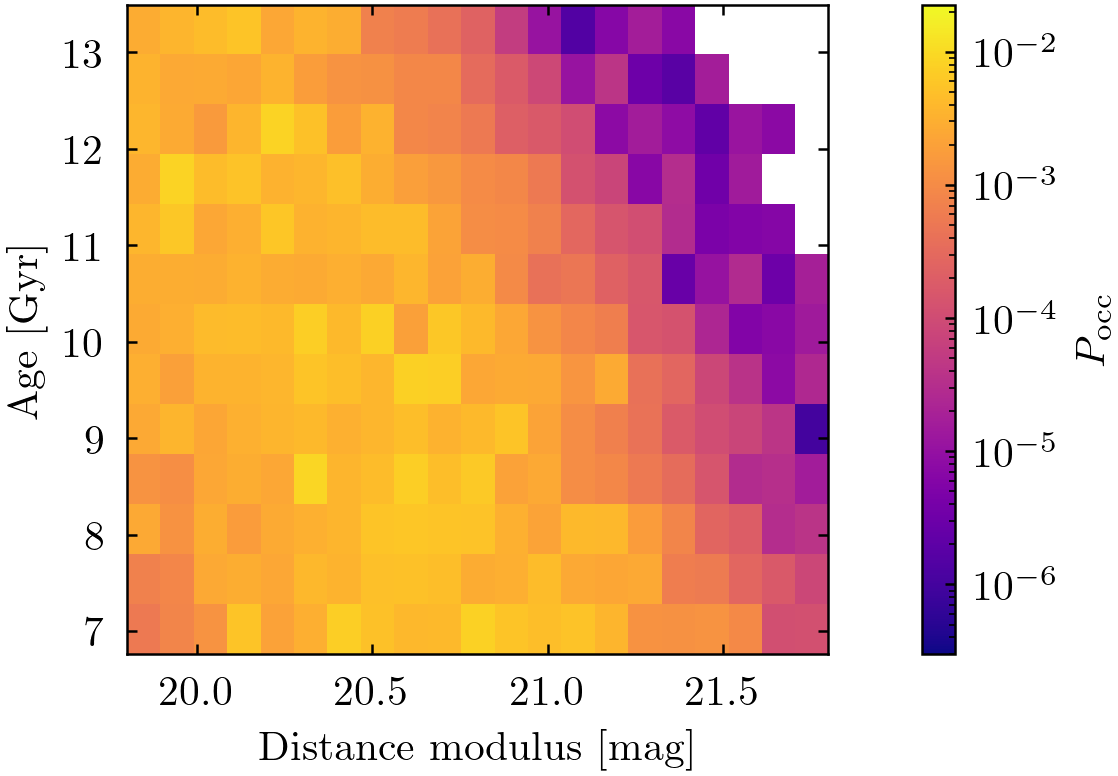}
    \caption{Occurrence probability of acceptable stochastic realisations in the forward-modelling analysis assuming a Kroupa IMF. Higher probabilities therefore indicate stellar populations that more naturally reproduce the observed properties of Fornax-7.}
    \label{fig:f7-prob}
\end{figure}

\begin{figure*}[h!]
    \centering
    \includegraphics[width=\linewidth]{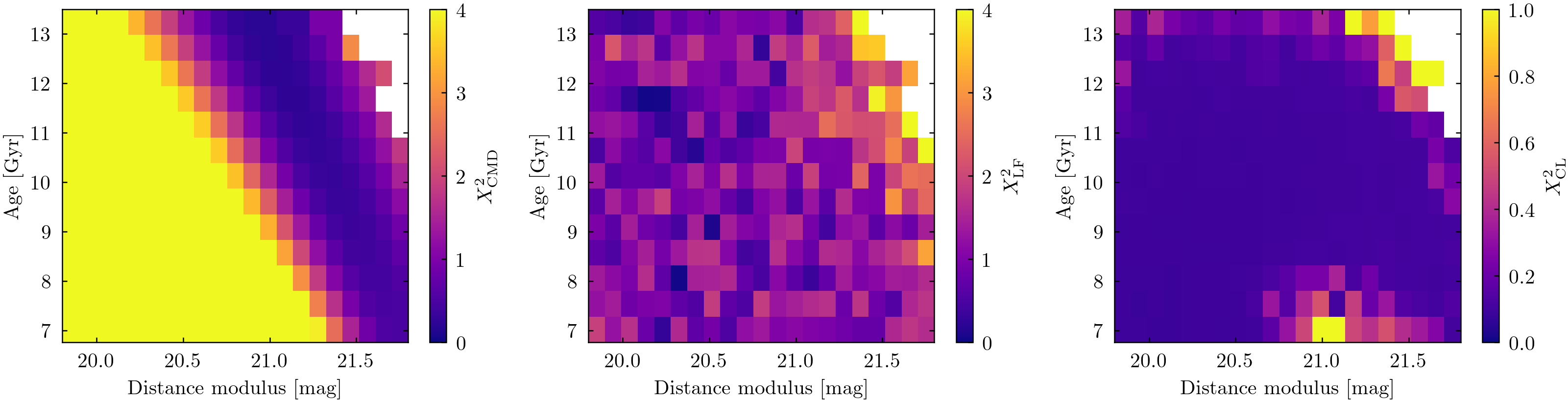}
    \caption{$\chi^2$ statistics for a given age and distance modulus, assuming a Kroupa IMF. The {\it left panel} shows $\chi^2_{\rm CMD}$ derived from the BEST-6. The {\it middle panel} presents the luminosity-function statistic $\chi^2_{\rm LF}$ based on the observed \IE-band LF of Fornax-7. The {\it right panel} shows $\chi^2_{\rm CL}$, the contribution from the integrated optical colours. Darker regions (lower values) indicate better agreement with the observations, whereas white regions (higher values) correspond to areas with no acceptable realisations.}
    \label{fig:f7-chi2}
\end{figure*}

As was described in the previous section, the analysis in Sect.~\ref{sc:analysis} was performed for two metallicity ranges: a strict range of $-2.2 \leq [{\rm M/H}] \leq -2.0$, motivated by ultra-faint dwarf galaxies, and a relaxed range of $-2.2 \leq [{\rm M/H}] \leq -1.0$, encompassing the old GCs of the Fornax dSph. We also considered both Kroupa and Chabrier IMFs. Unless stated otherwise, in this section, we adopt the results from the Kroupa IMF with the strict metallicity constraint as our fiducial model. Therefore, the figures shown here correspond to the Kroupa IMF with the stringent metallicity constraint. The corresponding figures for the Chabrier IMF are presented in Appendix\,\ref{app:chabrier}. Overall, comparing the results obtained with the Kroupa and Chabrier IMFs (Appendix\,\ref{app:chabrier}), we find excellent agreement between the inferred parameters. This demonstrates that the derived stellar population properties of Fornax-7 are robust against reasonable variations in the adopted IMF. Furthermore, we also present the results obtained using the Kroupa IMF with the relaxed metallicity constraint in Appendix~\ref{app:relaxed}.

The resulting $P_{\rm occ}$, which varies significantly across the age--distance parameter space, is presented in Fig.\,\ref{fig:f7-prob}. There is an overall gradient with increasing distance modulus, where the occurrence probability first increases and then decreases, exhibiting occurrence probabilities several orders of magnitude lower than the preferred regions. In addition, at lower distance moduli, older stellar populations are generally favoured, whereas at larger distance moduli, younger populations become more probable. The upper-right region of the parameter space, corresponding to both large distance moduli and old stellar populations, is strongly disfavoured. Furthermore, some parts of the parameter space yield $P_{\rm occ}=0$, indicating that no successful realizations satisfying constraints~1--3 were obtained. 

The corresponding $\chi^2$ distributions are shown in Fig.~\ref{fig:f7-chi2}. The most notable feature is the CMD $\chi^2$ (left panel), which exhibits a pronounced age--distance degeneracy, favours either an old stellar population located at the distance of the Fornax dSph or a younger population at a larger distance. 

\begin{figure}
    \centering
    \includegraphics[width=\linewidth]{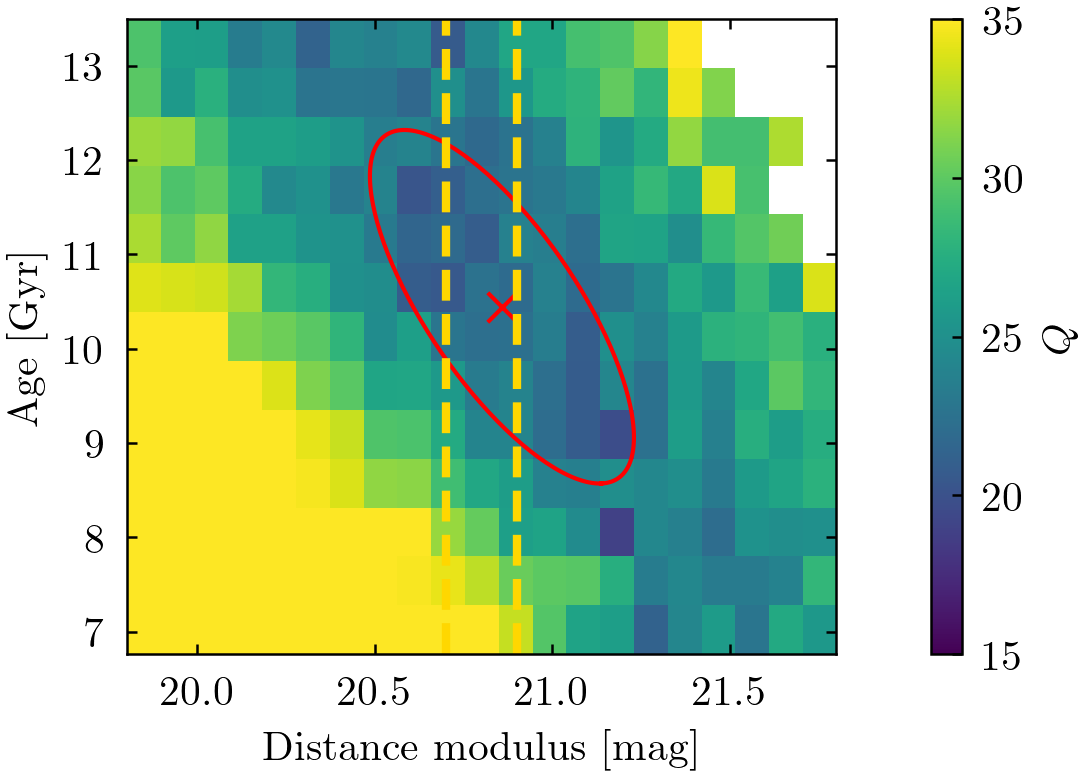}\\
    \vspace{6mm}
    \includegraphics[width=\linewidth]{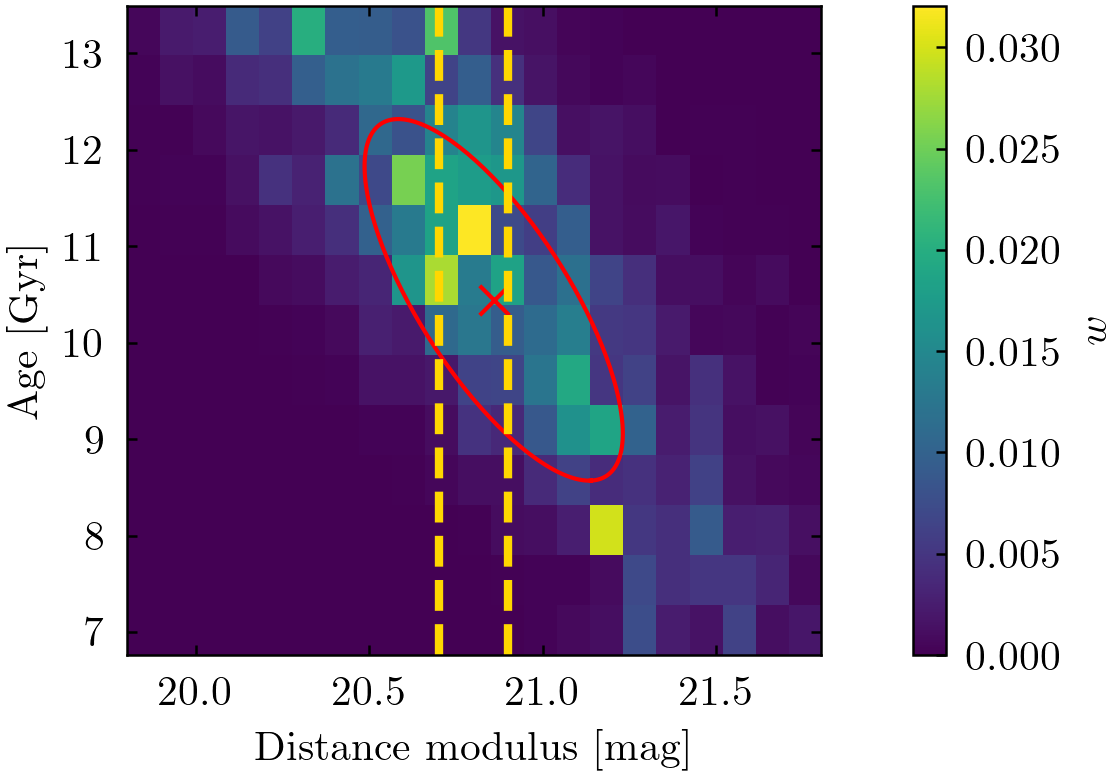}
    \caption{Distribution of the combined metric ($Q$) and the corresponding weights ($w$) in the age--distance modulus space assuming the Kroupa IMF. Lower values correspond to a better overall agreement with the observations. The red cross indicates the weighted mean solution while the red ellipse shows the covariance region derived from the weighted parameter distribution. The vertical dashed line marks the range of distance modulus of the Fornax dSph.}
    \label{fig:f7-metric}
\end{figure}

\begin{figure}
    \centering
    \includegraphics[width=\linewidth]{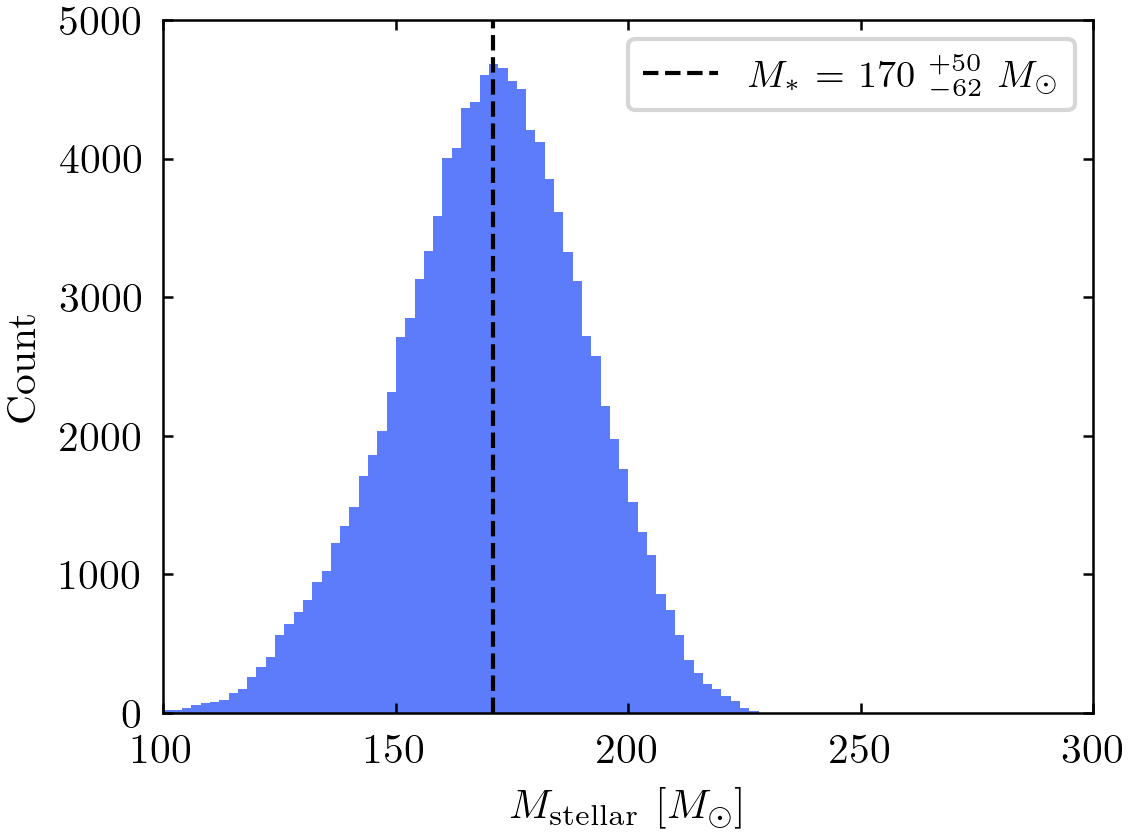}
    \caption{Posterior stellar-mass distributions of Fornax-7 derived from stochastic forward modelling within the covariance-selected parameter region, assuming a Kroupa IMF.}
\label{fig:stellar-mass}
\end{figure}

The combined metric $Q$
and the corresponding best-fitting parameters are presented in Fig.~\ref{fig:f7-metric}. The resulting stellar population parameters are summarised in Table~\ref{tab:final-params}. All explored configurations yield a distance modulus consistent with that of the Fornax dSph. The restrictive and relaxed metallicity analyses give $\mu \simeq 20.86 \pm 0.37$ and $20.82 \pm 0.33$ for Kroupa IMF, consistent with distance modulus of the Fornax dSph, $(m-M)_{\rm For}=20.8\pm 0.1$. This strongly suggests that Fornax-7 is located at the distance of the Fornax dSph. However, establishing whether the system is gravitationally associated with the galaxy will require follow-up radial-velocity measurements.

Under the restrictive metallicity prior, the best-fitting age is ($10.4 \pm 1.9$)\,Gyr, while allowing a broader metallicity range yields a slightly younger age of $(9.2 \pm 1.8)$\,Gyr. In both cases, Fornax-7 is clearly an old stellar system. The shift toward younger ages when more metal-rich solutions are permitted reflects the well-known age--metallicity degeneracy. Future follow-up observations, either through deeper photometry or spectroscopy, will be required to break this degeneracy and further constrain the stellar population properties of the system. 

For the relaxed metallicity range, we find a best-fitting value of $[{\rm M/H}]\simeq-1.4\pm0.3$. We emphasize that constraining the metallicity is not a primary goal of this analysis; rather, the broader metallicity range is explored to assess the robustness of the inferred age and distance modulus. With the current data, we do not expect the metallicity itself to be well constrained, particularly at $[{\rm M/H}]\lesssim-1.5$, where the differences between isochrones of different metallicities become increasingly small. Although other stellar-evolution models, such as BaSTI (\citealp{basti2021}), extend to lower metallicities, at this level the systematic differences between different sets of stellar models may become comparable to, or larger than, the differences induced by metallicity. We therefore do not extend the present analysis to additional isochrone sets, leaving a more detailed investigation of the metallicity to future work with deeper or complementary follow-up observations.

The stellar mass of Fornax-7 was estimated from the subset of stochastic realisations lying within the covariance region of the preferred parameter space. The stellar-mass distribution of these accepted realisations was then used to determine the most likely stellar mass and its uncertainty. The resulting distribution is shown in Fig.~\ref{fig:stellar-mass}. We find a stellar mass of $170^{+50}_{-62}\,M_{\odot}$ for the Kroupa IMF, while we obtain a stellar mass of only a few per cents lower for the Chabrier IMF. The inferred stellar mass is essentially unchanged when adopting the relaxed metallicity prior, and we find $M_\star =163^{+50}_{-49}\,M_{\odot}$. Overall, the agreement between all explored models indicates that Fornax-7 contains only a few hundred solar masses in stars, making it one of the least massive known old stellar systems. From our measured integrated magnitude of $\IE=20.8\pm0.2$ and the expected colour transformation for an old, metal-poor stellar population (\citealp[Appendix A]{saifollahi2025a}), corresponding to $V - \IE=0.3\pm0.1$, we derive an absolute magnitude of $M_V = +0.3 \pm 0.3$, where the uncertainty includes both the photometric measurement and the colour transformation. These parameters, together with the other derived properties of Fornax-7, are summarised in Table~\ref{tab:properties}.

\begin{figure}[h!]
    \centering
    \ \ \ \ \ \ \ \ \ \ \ \ Luminosity-Size relation
    \\[2pt]
    \includegraphics[width=\linewidth]{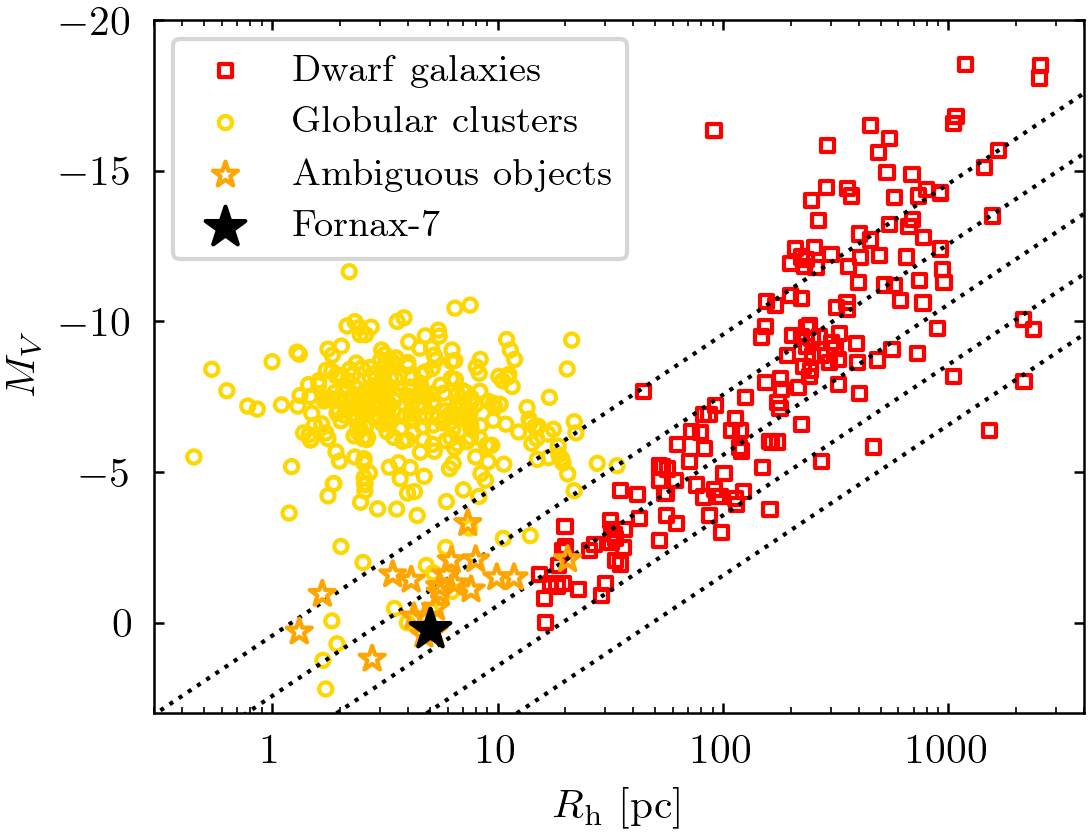}
    \\[5pt]
    \ \ \ \ \ \ \ \ \ \ \ \ Metallictiy-Luminosity relation
    \\[2pt]
    \includegraphics[width=\linewidth]{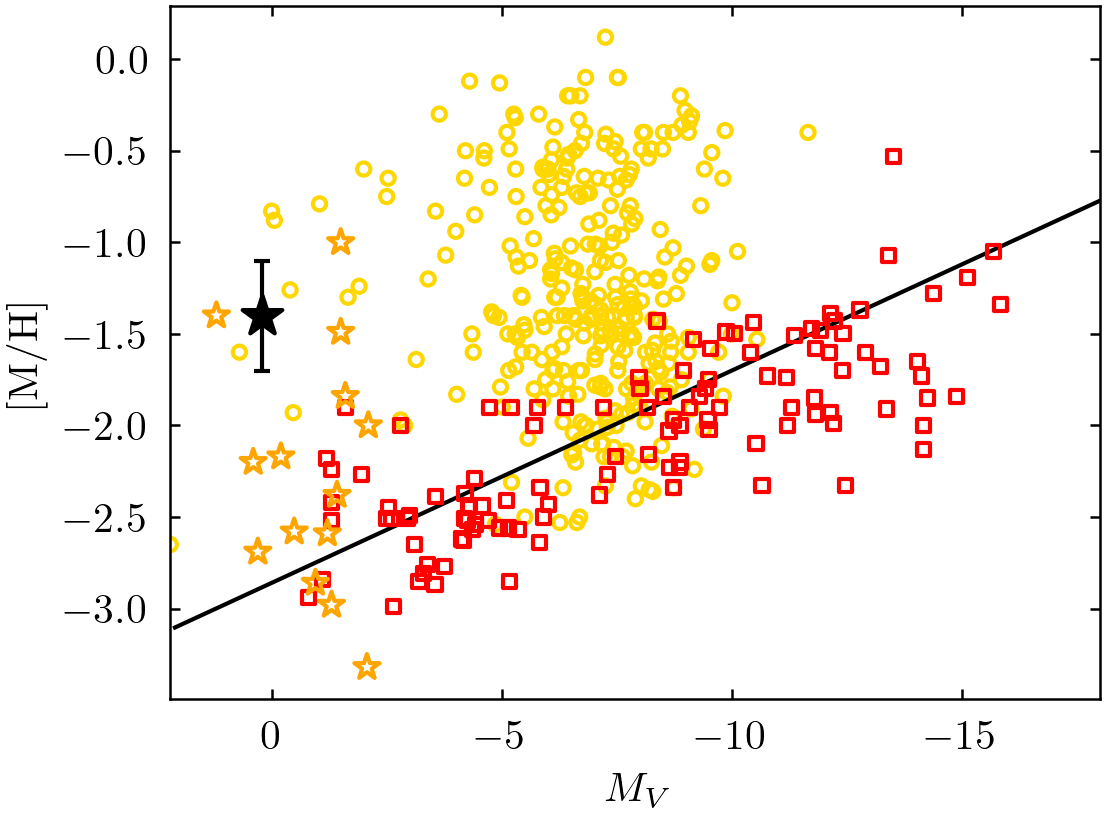}
    \caption{Comparison of Fornax-7 properties (black star) against dwarf galaxies (red squares), GCs (yellow circles), and ambiguous stellar systems (orange stars). \textit{Top}: Luminosity--size relation, showing the absolute $V$-band magnitude as a function of projected half-light radius. The dotted lines indicate constant mean surface brightness levels of 24, 26, 28, 30, and 32~mag~arcsec$^{-2}$. \textit{Bottom}: Metallicity--luminosity relation. The solid black line shows the luminosity--metallicity relation for dwarf galaxies from \citet{ufd}. The data are from \citet{pace2025}, v1.1.1. The plots were generated using the notebook provided at \url{https://github.com/apace7/local_volume_database/tree/main}, with minor modifications.}
\label{fig:scaling-relations}
\end{figure}

Our results show that Fornax-7 is among the faintest and smallest systems ever identified. In the luminosity--size plane, the structural properties of Fornax-7 place it among the population of ambiguous stellar systems (Fig.~\ref{fig:scaling-relations} top panel), lying very close to the sequence by dwarf galaxies. In the luminosity--metallicity plane (Fig.~\ref{fig:scaling-relations} bottom panel), Fornax-7 occupies a region of parameter space populated by metal-poor GCs and other ambiguous stellar systems, while exhibiting a higher metallicity than expected for a dwarf galaxy of its luminosity. Instead, it occupies a region of parameter space more consistent with metal-poor GCs and other ambiguous stellar systems. However, the quality of the current data does not allow us to place robust constraints on the metallicity. While we briefly consider this in the discussion, more definitive constraints will require follow-up observations. However, this comparison should be interpreted with caution, as the PARSEC isochrones used in our forward modelling extend only down to $[{\rm M/H}]=-2.2$. If Fornax-7 is intrinsically more metal-poor, as suggested by an extrapolation of the luminosity--metallicity relation of \citet{ufd}, our analysis would not be able to recover such lower metallicities. Regardless, the data are unlikely to provide sufficiently precise constraints on the metallicity, and we primarily investigate this possibility as a consistency check for the distance modulus and the inferred age.

Currently, a small number of ultra-faint stellar systems with similar or even lower stellar masses than Fornax-7 are known. Examples include Segue~3 (\citealp{Belokurov2010}), UNIONS~I/Ursa Major~III (\citealp{smith2024}), and the star cluster associated with Ursa Major~II (\citealp{Eadie2022}). These systems are old and metal-poor, but they have been suggested to be the products of tidal interactions with their host galaxies, representing the low-mass remnants of initially more massive systems (\citealp{Fadely2011,Eadie2022,errani2024}). Moreover, Fornax-7 is located at a projected distance of at least 2.8~kpc from the centre of the Fornax dSph (assuming a distance of 145~kpc), placing it in the outer regions of the galaxy. This separation is comparable to the estimated tidal radius of the Fornax dSph, $\sim3$~kpc (\citealp{des-wang-2019}), suggesting that the system, if bound to Fornax, has likely evolved in a relatively weak tidal environment, although this interpretation remains uncertain.

\begin{table*}
\caption{Best-fitting stellar population parameters and stellar masses obtained from the forward-modelling analysis, adopting the strict (${-2.2 < \rm [M/H}] \leq -2.0$) and relaxed (${-2.2 \leq \rm [M/H}] \leq -1.0$) metallicity constraints.}
\label{tab:final-params}
\centering
\begin{tabular}{llcccc}
\hline
 & IMF &
Age [Gyr] &
$[{\rm M/H}]$ &
$\mu$ [mag] &
$M_{\rm stellar}$ [$M_\odot$] \\
\hline
Strict metallicities (${-2.2 \leq \rm [M/H}] \leq -2.0$)
& Kroupa
& $10.4 \pm 1.9$
& $...$
& $20.86 \pm 0.37$
& $170^{+50}_{-62}$ \\ 
\vspace{0.5\baselineskip}
& Chabrier
& $10.5 \pm 1.8$
& $...$
& $20.93 \pm 0.37$
& $165^{+54}_{-59}$ \\
\hline
Relaxed metallicities (${-2.2 \leq \rm [M/H}] \leq -1.0$)
& Kroupa
& $9.2 \pm 1.8$
& $-1.4 \pm 0.3$
& $20.82 \pm 0.33$
& $163^{+50}_{-49}$ \\
\hline
\end{tabular}
\end{table*}

\begin{table}
\centering
\caption{Derived properties of Fornax-7.}
\label{tab:properties}
\begin{tabular}{ll}
\hline
Property & Value \\
\hline
\vspace{0.5\baselineskip}
Right Ascension (J2000) & \ra{02;35;17.94} \\ 
\vspace{0.5\baselineskip}
Declination (J2000) & \ang{-33;57;24.25}
\\
\vspace{0.5\baselineskip}
Projected distance from Fornax dSph & $2.8$\,kpc \\
\vspace{0.5\baselineskip}
\IE Apparent magnitude, $M_{\IE}$ & $20.8\pm0.2$ \\
\vspace{0.5\baselineskip}
\IE Absolute magnitude, $M_{\IE}$ & $0.0\pm0.2$ \\
\vspace{0.5\baselineskip}
{\it V}-band Absolute magnitude, $M_V$ & $+0.3\pm0.3$ \\
\vspace{0.5\baselineskip}
$g-r$ & $0.3 \pm 0.1$ \\
\vspace{0.5\baselineskip}
$g-i$ & $0.45 \pm 0.1$ \\
\vspace{0.5\baselineskip}
Half-light radius, $R_{\rm h}$ & $(5.0\pm1.0)$\,pc \\
\vspace{0.5\baselineskip}
Effective surface brightness, $\langle\mu\rangle_{{\rm e},\IE}$ & $26.50$\,mag\,arcsec$^{-2}$ \\
\vspace{0.5\baselineskip}
Effective surface brightness, $\langle\mu\rangle_{{\rm e},V}$ & $26.80$\,mag\,arcsec$^{-2}$ \\
\vspace{0.5\baselineskip}
Ellipticity & $0.02^{+0.35}_{-0.02}$\\
\vspace{0.5\baselineskip}
Position angle & $116^{+22}_{-30}$\,deg\\
\vspace{0.5\baselineskip}
Stellar mass & $170^{+50}_{-62}~M_\odot$ \\
\hline
\end{tabular}
\end{table}

\section{\label{sc:dis}  Discussion: The nature of Fornax-7}

Based on the estimated properties of Fornax-7, there are a few possible scenarios for the nature of Fornax-7. If associated with the Fornax dSph, Fornax-7 could represent an extremely low-mass and remote star cluster. Such a system would provide a rare opportunity to study star cluster formation, survival, and the stellar mass function at the lowest GC masses. This could be relevant to the long-standing debate regarding the dark matter distribution and the GC timing-problem of the Fornax dSph (\citealp{fornax3,fornax4,fornax1,fornax2,genina2022}). Alternatively, Fornax-7 could represent a low-mass dwarf satellite galaxy of Fornax dSph, making it the first known example of a {``satellite of a satellite''} around a host galaxy as low in mass as Fornax. 

The distinction between the two scenarios can only be directly tested through measurements of the internal velocity dispersion. Given the faintness of the targets, it cannot be done with current facilities and such measurements will likely require the next generation of extremely large telescopes and their instruments, such as the Extremely Large Telescope (ELT), as high-resolution spectroscopy of stars this faint is beyond the capabilities of current facilities. A potential observational discriminator between the star cluster and dwarf galaxy scenarios is the degree of mass segregation. As shown by \citet{errani2025}, dynamically relaxed star clusters are expected to exhibit significant mass segregation, whereas dark matter-dominated systems should preserve a much weaker stellar mass dependence in their spatial distribution. The current \Euclid observations are not sufficiently deep to probe the stellar mass range required for this test. However, future deep imaging with facilities such as the HST and JWST could potentially measure the radial distributions of low- and high-mass stars, providing an additional constraint on the nature of Fornax-7. In the meantime, dwarf galaxies have been shown to exhibit significantly larger internal metallicity spreads than GCs, especially at the faint regime (\citealp{willman2012}), providing another potential way to assess the nature of Fornax-7 in the future .

Here, we discuss what can be inferred from the present observations and the expected dynamical evolution of Fornax-7 under each of the proposed scenarios.

\subsection*{The environment of Fornax-7}

Fornax-7 is located at a projected distance of 2.8~kpc from the centre of the Fornax dSph, implying a three-dimensional separation of at least 2.8~kpc. Assuming a circular orbit, this suggests that Fornax-7 has likely spent most of its lifetime in the outer regions of its host, where the tidal field  is relatively weak. We therefore investigate whether its dynamical evolution is expected to have been significantly affected by the gravitational potential of the Fornax dSph. Following \cite{errani2025}\footnote{Adopted from \citet{Spitzer1987}.}, the internal crossing time of a stellar
system can be estimated as
\begin{equation}
T_{\rm cross}
=
R_{\rm h}^{3/2}\left(GM_{\rm h}\right)^{-1/2}
=
R_{\rm h}^{3/2}
\left(GM_\star\Upsilon_{\rm dyn}\right)^{-1/2}\,,
\end{equation}
where $M_{\rm h}$ is the dynamical
mass of the system within $R_{\rm h}$ (half-light radius), $M_\star$ is its stellar mass within $R_{\rm h}$, and
$\Upsilon_{\rm dyn}\equiv M_{\rm h}/M_\star$ is the dynamical-to-stellar
mass ratio. Under the star-cluster hypothesis, we adopt
$\Upsilon_{\rm dyn}=1$, corresponding to a system without a
dynamically significant dark matter component. Using
$M_\star=85\,M_\odot$ and $R_{\rm h}=5.0$~pc, we obtain $T_{\rm cross}\simeq18.1~{\rm Myr}$. This value represents an upper limit, as the crossing time would be shorter if Fornax-7 were embedded in a dark matter halo. For example, assuming a halo mass of $\sim10^7\,M_\odot$ or $\sim10^8\,M_\odot$, would further reduce $t_{\rm cross}$. Furthermore, the orbital period of Fornax-7 around the Fornax dSph can be approximated as
\begin{equation}
T_{\rm orb}
=
2\pi
\sqrt{\frac{{R_0}^3}{G\,M(<R_0)}}\,,
\end{equation}
where $R_0$ is the orbital radius and $M(<R_0)$ is the mass enclosed within that radius. Assuming a de-projected orbital radius of $R_0=2.8$~kpc (a lower limit) and an enclosed mass of $M(<R_0)\simeq 4 \times 10^8\,M_\odot$ (\citealp{Kowalczyk2019}), we obtain an orbital period of $T_{\rm orb}\simeq690\,{\rm Myr}$. Comparing this with the internal crossing time, $T_{\rm cross}\simeq18.1$\,Myr, yields ${T_{\rm orb}}/{T_{\rm cross}}\simeq38$. This indicates that the internal crossing time is much shorter than orbital time, implying that if there is any perturbation to the cluster, it recovers more quickly to a close-to-equilibrium state than the time to complete an orbit. Furthermore, we estimate the present-day tidal radius of Fornax-7. Using the orbital radius $R_0 = 2.8$\,kpc and the mass $M_{\mathrm{cl}}$, we calculate the theoretical tidal radius of the star cluster as derived by \cite{Bertin08}\footnote{Re-derived from \citealp{Spitzer1987}.} for a spherically symmetric host galaxy and a Keplerian cluster-potential as
\begin{equation}
   R_{\mathrm{t}}^{\mathrm{cl}}=\left(\frac{GM_{\mathrm{cl}}}{\Omega^2 \upsilon}\right)^{1/3}\,,
   \label{eq2}
\end{equation}
where the orbital frequency $\Omega$ at $R_0$ in a circular orbit, the epicyclic frequency $\kappa$ at $R_0$, and a positive dimensionless coefficient $\upsilon$ are defined as
\begin{align}
\Omega^2 &= \left(\frac{\mathrm{d}\Phi_{\mathrm{For}}(r)}
{\mathrm{d}r}\right)_{R_0}/R_0\,, &&\\
\kappa^2 &= 3\Omega^2 +
\left(\frac{\mathrm{d}^2\Phi_{\mathrm{For}}(r)}
{\mathrm{d}r^2}\right)_{R_0}\,, &&\\
\upsilon &= 4-\frac{\kappa^2}{\Omega^2}\,. &&
\end{align}

Assuming an enclosed dark matter halo mass of $4 \times 10^8\,M_\odot$ (NFW profile, \citealp{nfw}) for the Fornax dwarf and a mass of $170\,M_\odot$ for Fornax-7, the resulting tidal radius is approximately $20.5\,\mathrm{pc}$. This corresponds to a ratio of ${R_{\rm h}}/{R_{\mathrm{t}}^{\mathrm{cl}}}
\simeq {5.0}/{20.5} \simeq
0.25$, indicating that the observed stellar distribution lies well within the tidal radius. Therefore, the present-day structure of Fornax-7 is not expected to be strongly affected by the tidal field of the Fornax dSph. This would suggest that Fornax-7's observed stellar distribution is expected to be largely governed by its own self-gravity rather than by ongoing tidal stripping. These conclusions should be regarded as tentative, since the orbital properties and dynamical history of Fornax-7 are currently unknown. They are based on simple analytical estimates, including the comparison between the internal crossing time, orbital period, and the present-day tidal radius, but do not account for the system’s orbital evolution or possible past tidal interactions.

\subsection*{Scenario 1 -- remote star cluster of the Fornax dSph}

If Fornax-7 is a star cluster (scenario~1), given the argument above, its long-term evolution is expected to be governed primarily by dynamical evaporation through two-body relaxation. Consequently, its present-day stellar mass of $\sim170\,M_\odot$ cannot represent its initial mass. 

A star cluster with the present-day properties of Fornax-7 would be expected to dissolve completely within a few Gyr. We estimate Fornax-7's relaxation time using equation~(5)
of \citet{errani2025},
\begin{equation}
T_{\rm rel} \approx
\frac{1}{\sqrt{12\pi G}}
\left(\Upsilon_{\rm dyn} R_{\rm h}\right)^{3/2}
N_\star M_\star^{-1/2}
\left[\ln(\Lambda)-1.9\right]^{-1}\,,
\end{equation}
where $N_\star$ is the number of stars within $R_{\rm h}$. Again, we assume
$\Upsilon_{\rm dyn}=1$, appropriate for a purely stellar system,
and adopt $\ln(\Lambda)=8.2$, following the footnote of
\citet{errani2025}. Assuming a mean stellar mass of
$\bar{m}_\star=0.25\,M_\odot$, the present-day stellar mass of
Fornax-7, $M_\star\simeq85\,M_\odot$, corresponds to
$N_\star\simeq340$. For the observed half-light radius of
$R_{\rm h}=5.0$~pc, we obtain a relaxation time of $T_{\rm rel}\simeq160\,{\rm Myr}$. Then, following
\citet{putten2012}, we approximate the evaporation time of an isolated stellar system as $T_{\rm ev}\simeq20.5\,T_{\rm relax} \simeq3.28~{\rm Gyr}$.

If Fornax-7 is a purely stellar system, it would have undergone significant dynamical evolution and mass loss over its lifetime. We use the present-day mass of Fornax-7 to estimate its initial mass based on an assumed evaporation timescale of $13$~Gyr for the Fornax-7 progenitor. This time corresponds to the current age of Fornax-7 ($\sim10$~Gyr) plus its estimated remaining lifetime ($\sim3$~Gyr, as estimated for $T_{\rm ev}$ of Fornax-7). Using the same formalism, and assuming that Fornax-7 half-light radius has remained mostly unchanged, this gives an initial stellar mass of approximately $10^4\,M_\odot$. Assuming that Fornax-7 was initially more compact, its initial stellar mass could have been a few times higher. 

The next question is whether such a cluster could have remained in the outskirts of the Fornax dSph throughout its evolution. Previous observations and dynamical modelling of the Fornax GC system suggest that remote star clusters, in particular Fornax-1, Fornax-2, and Fornax-5, have likely spent most of their lifetimes at large galactocentric distances, experiencing only modest orbital variations (\citealp{Borukhovetskaya2022}). For instance, Fornax-1 with $M_\star \approx 4.2\times10^4\,M_\odot$ (\citealp{deboer2016}), located at a smaller projected galactocentric distance ($\sim1.6$~kpc), is not expected to sink to the centre of the Fornax dSph within 10~Gyr. Given its much lower mass and larger projected distance ($\sim2.8$~kpc), dynamical friction should be even less effective on Fornax-7. This suggests that Fornax-7 is unlikely to suffer from the timing problem and it is possible that long-lived star clusters remain in the outer halo of the Fornax dSph without experiencing significant orbital decay due to dynamical friction. Therefore, from both observational and theoretical perspectives, the scenario in which Fornax-7 is a remote GC associated with the Fornax dSph remains plausible. 

\subsection*{Scenario 2 -- satellite of satellite}

If Fornax-7 is a dwarf galaxy associated with the Fornax dSph, abundance-matching relations predict that it should reside in a dark matter halo with a mass of order $\sim10^8\,M_\odot$. We adopt a halo mass for Fornax-7 using the stellar-to-halo mass relation (SHMR) of \citet{moster2013}, extrapolated to the ultra-low stellar mass regime (see Fig.\,2 in \citealp{sales-nat}). Although the scatter in the SHMR is expected to increase substantially at these masses (\citealp{munshi2021,kim2026}), the inferred halo mass should be regarded as an order-of-magnitude estimate. Nevertheless, \citet{santos-santos-2022}, examining a number of simulations, suggest that, if the SHMR continues as a steep power law towards lower masses, the resulting relation remains broadly consistent with the extrapolation of \citet{moster2013}. This motivates our adoption of the extrapolated SHMR as a fiducial estimate. We note, however, that substantially lower halo masses, potentially as low as $\sim10^7\,M_\odot$, are also possible (\citealp{Kravtsov2022,brown2016}).

The existence of a $\sim10^7$--$10^8\,M_\odot$ satellite around a dwarf galaxy is not unexpected within the $\Lambda$CDM framework. Depending on the peak halo mass of the host dwarf galaxy, it may or may not be expected to host a subhalo with a mass of $\sim10^8\,M_\odot$ (see Fig.~7 of \citealp{santos-santos-2022} and Fig.\,1 of \citealp{Vitral2025}). If the Fornax dSph reached a peak halo mass of $\sim10^{10}\,M_\odot$, the presence of such a subhalo is expected. In contrast, if its peak halo mass was only $\sim10^9\,M_\odot$, the expected number of $10^8\,M_\odot$ subhaloes is essentially zero. On the other hand, if Fornax-7 resides in a lower-mass dark matter halo of $\sim10^7\,M_\odot$, its presence would be expected even for a lower-mass Fornax halo. Furthermore, high-resolution simulations from the EDGE project predict that dwarf galaxies with masses comparable to the Fornax dSph, with halo masses between $10^9$ and $10^{10}\,M_\odot$, typically host between zero and two subhaloes of order of $\sim10^8\,M_\odot$ (\citealp{ethan2025}). 

Such a massive subhalo would experience significantly stronger dynamical friction than a purely stellar system (a star cluster), potentially causing it to spiral towards the centre of the Fornax dSph on a timescale of a few Gyr. At first, this appears difficult to reconcile with the present projected separation of at least 2.8\,kpc. However, several mechanisms could alleviate this apparent tension. Numerical studies have shown that interactions between the Fornax dSph and the Milky Way can dynamically heat both the stellar and dark matter components of the Fornax dSph and may scatter satellites and GCs of the Fornax dSph onto larger orbits. Simultaneously, these interactions affect the internal structure of the satellite galaxies (here Fornax dSph) through tidal heating, however, the observational evidence remains mixed (\citealp{Battaglia2015}).

\citet{Vitral2025} argued that the passage of massive subhaloes through the Fornax dSph can redistribute its stellar component. If Fornax-7 is indeed a dark matter-dominated satellite, it could therefore represent one such perturber. In another scenario, dwarf--dwarf interactions involving the Fornax dSph and a relatively massive dwarf galaxy (with a mass ratio of $\sim$1:2--1:5) have been proposed to explain the observed properties of the Fornax dSph (\citealp{Leung2020}), as well as the complex star formation history of Fornax-1 and Fornax-2 GC systems (\citealp{Rusakov2021}). However, Fornax-7 does not appear to be a plausible candidate for such a companion. Even under the assumption of a total halo mass of $\sim10^8\,M_\odot$, it would still be an order of magnitude less massive than the present-day Fornax dSph. One possibility is that Fornax-7 represents the surviving remnant of a substantially more massive system involved in such a past interaction.

Moreover, the Fornax dSph was likely significantly more massive in the past prior to tidal stripping by the Milky Way (\citealp{read2019}), making such a major merger scenario even less likely. Nevertheless, a past merger involving the Fornax dSph could still provide a possible explanation for the large galactocentric distance of Fornax-7. Such an interaction could perturb pre-existing or recently formed GCs onto wider orbits, potentially increasing their survival times against dynamical friction (\citealp{Leung2020}). Indeed, a past merger has previously been invoked to explain the metallicity distribution of the Fornax GCs (\citealp{Larsen2012}), while independent evidence for a merger history has been reported (\citealp{Amorisco2012,Piatti2014,delPino2017}). Alternatively, Fornax-7 could be a recently accreted system by the Fornax dSph after previously orbiting within the Milky Way halo.

Furthermore, If future spectroscopic observations confirm that Fornax-7 is a UFD associated with the Fornax dSph, its existence could provide valuable constraints on galaxy formation at the lowest halo masses. \citet{ahvazi2024} show that the luminous occupation fraction of haloes drops rapidly in this regime, reaching 50\% at $M_{\rm peak}\sim9\times10^7,M_\odot$, with different models predicting substantially different transitions between luminous and dark haloes. In particular, \citet{santos-santos-2022} show that models invoking a low-mass cutoff in galaxy formation and those in which the stellar mass–halo mass relation continues as a steep power law predict markedly different populations of ultra-faint satellites around dwarf galaxies. With a stellar mass of only about $170,M_\odot$, Fornax-7 lies in this extreme regime and could therefore provide a useful observational constraint on the transition between luminous and dark haloes.

\subsection*{Scenario 3? -- UFD in outer halo of the Milky Way}

Our analysis strongly suggests that Fornax-7 lies at the distance of the Fornax dSph, but its dynamical association with the galaxy remains to be confirmed. Follow-up spectroscopic observations are required to measure its radial velocity and test whether the system is gravitationally bound to the Fornax dSph. If future observations demonstrate that Fornax-7 is not dynamically associated with the Fornax dSph, it would instead represent one of the most distant and lowest-mass UFD candidates currently known in the outer halo of the Milky Way. Such an object would provide valuable constraints on the minimum stellar and halo mass capable of forming and sustaining a galaxy in the absence of significant mass loss through tidal interactions, and hence on the limits of galaxy formation, and on reionisation feedback.

\section{\label{sc:summary} Summary}

We have presented the analysis of Fornax-7, an ultra-faint stellar system identified in the Euclid Data Release (DR1) in the direction of the Fornax dSph. Using the deep \Euclid \IE imaging together with complementary optical and near-infrared data, we measured its structural parameters and investigated its stellar population through extensive forward modelling based on stochastic stellar population realisations.

The structural analysis yields a projected half-light radius of $R_{\rm h}=(5.0\pm1.0)\,{\rm pc}$, under the hypothesis that the system is at the distance of the Fornax dSph. This makes Fornax-7 one of the smallest known ultra-faint stellar systems. Forward modelling performed using both Kroupa and Chabrier IMFs produces highly consistent results, demonstrating that the inferred parameters are largely insensitive to the adopted IMF. We find that Fornax-7 hosts an old ($\sim$9--10\,Gyr), most likely metal-poor stellar population with a stellar mass of only $\sim170^{+50}_{-62}\,M_\odot$, and a distance modulus $\mu = 20.86 \pm 0.37$, consistent with that of the Fornax dSph. These derived properties place Fornax-7 among the faintest, least massive, and most compact old stellar systems known. In the luminosity--size and luminosity--metallicity planes, it occupies the region of parameter space shared by GCs, the faintest dwarf galaxies, and other ambiguous stellar systems.

The current data do not uniquely determine the nature of Fornax-7. If dynamically associated with the Fornax dSph, it could represent either one of its most remote and least massive GCs or an ultra-faint dwarf galaxy, making it the first known satellite of such a low-mass galaxy. Conversely, if future spectroscopy shows that Fornax-7 is not associated with the Fornax dSph, it would instead represent one of the most distant and lowest-mass ultra-faint dwarf galaxy candidates in the outer halo of the Milky Way, providing valuable constraints on the minimum halo mass capable of forming a galaxy in the absence of significant tidal stripping. Once Fornax-7 properties are be more firmly established by additional observations, a set of properly suited $N$-body simulations would be very useful to explore the evolutionary path of the system (Appendix\,D in \citealp{bellazini2026}). Our results and the discussion in this paper demonstrate the wide range of scientific questions that the discovery and future characterisation of Fornax-7 can address.

\begin{acknowledgements}

We would like to thank Christian Boily and Raphaël Errani for helpful discussions on dynamics of low mass stellar systems. We also thank Annus Haider for providing their synthetic photometry catalogue of stochastic stellar population models extending to lower stellar masses, which we used to perform consistency checks on our models. TS and PB acknowledge funding from the CNES postdoctoral fellowship programme. TS and AL acknowledge support from the Interdisciplinary Thematic Institute IRMIA++, as part of the ITI 2021–2028 program of the University of Strasbourg, CNRS, and Inserm, supported by IdEx Unistra (ANR-10-IDEX-0002) and the SFRI-STRAT’US project (ANR-20-SFRI-0012) under the framework of the French Investments for the Future Program. AMNF is supported by UK Research and Innovation (UKRI) under the UK government’s Horizon Europe funding guarantee [grant number EP/Z534353/1] and by the UK Science and Technology Facilities Council [grant number ST/Y001281/1].

\AckDRone
\AckEC  
\end{acknowledgements}

\bibliography{Euclid, Euclid-2, DR1, mybib} 

\begin{appendix}
\onecolumn 

\section{Fornax-7 in ground-based data}

\begin{figure*}[h]
    \centering
    \includegraphics[width=0.83\linewidth]{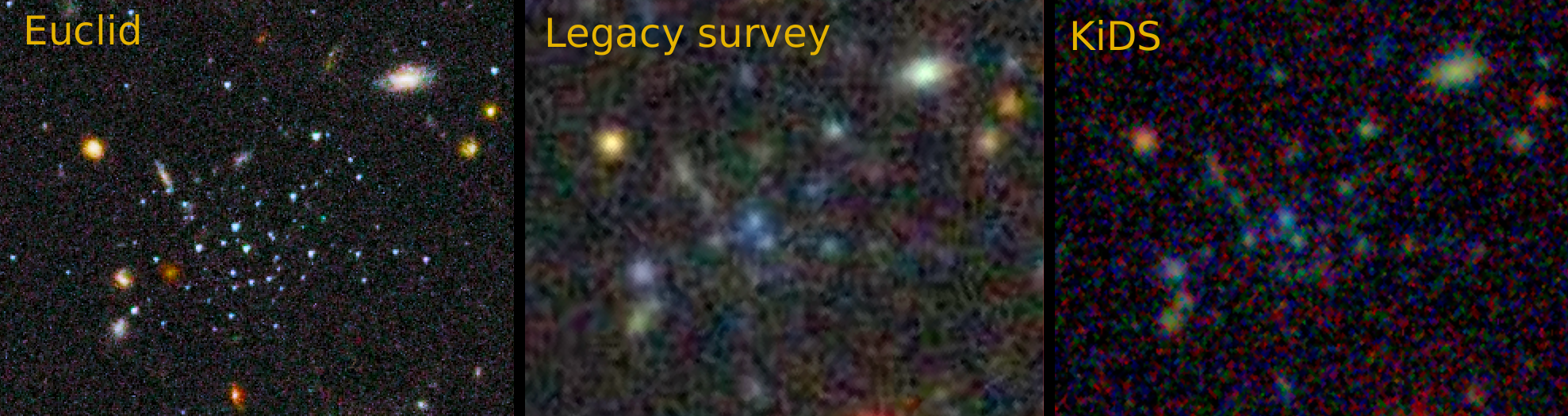}
    \caption{Colour images of Fornax-7 from \Euclid, the Legacy Survey, and the KiDS survey. The last two were retrieved and visualised using the Aladin Sky Atlas (\citealp{aladin1,aladin2}).}
    \label{euclid-vs-ground}
\end{figure*}

Here, in Fig.\,\ref{euclid-vs-ground} we compare the view of Fornax-7 in \Euclid with that from two ground-based surveys, the Legacy Survey and KiDS. The BEST-6 stars are also visible in the ground-based data, while the stellar overdensity associated with Fornax-7 becomes apparent only in the deeper and higher-resolution \Euclid imaging.

\section{Fornax-7 CMD in other bands}

\begin{figure*}[h]
    \centering
    \includegraphics[width=0.3\linewidth]{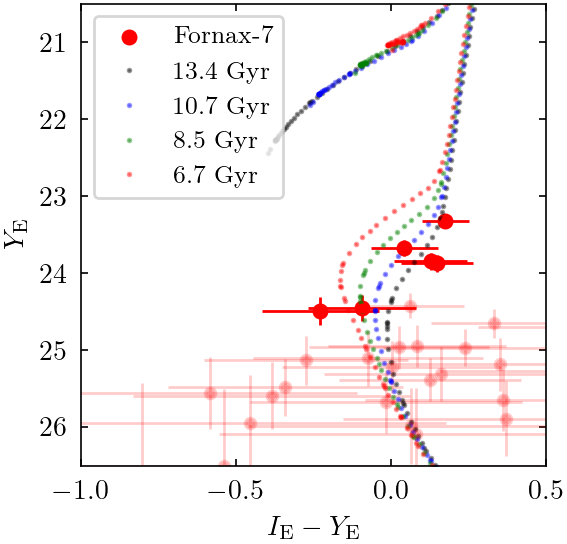}
    \includegraphics[width=0.3\linewidth]{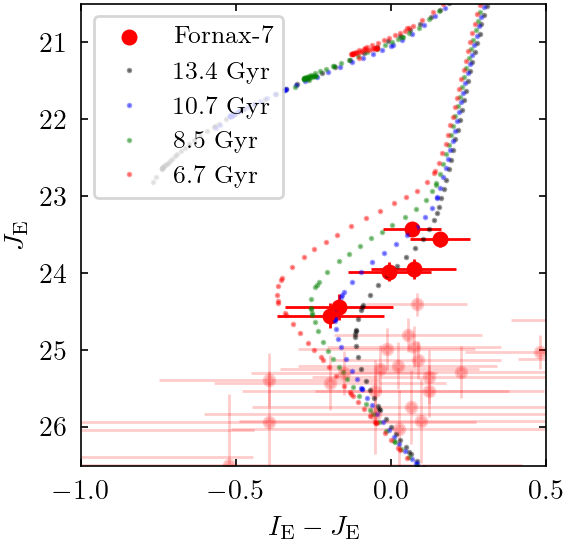}
    \includegraphics[width=0.3\linewidth]{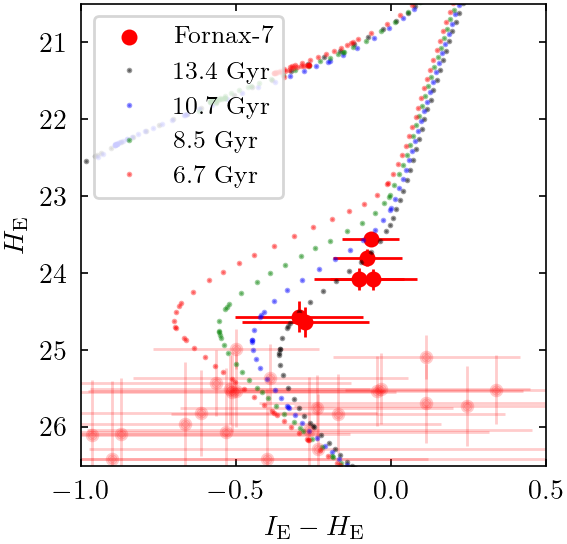}
    \caption{CMD of the Fornax-7 stars in the $(\IE-\YE,\YE)$, $(\IE-\JE,\JE)$, and $(\IE-\HE,\HE)$ planes. The larger red points show the selected stars used in the CMD analysis (BEST-6), while the coloured curves represent PARSEC isochrones with $[{\rm M/H}] = -2.2$ and distance modulus $m-M = 20.8$ for stellar populations with different ages.}
    \label{fig:f7-cmd-extra}
\end{figure*}

In addition to the main $(\IE-\HE,\HE)$ CMD presented in the main text, we also examined the Fornax-7 stars in the $(\IE-\YE,\YE)$ and $(\IE-\JE,\JE)$ spaces (see Fig.\,\ref{fig:f7-cmd-extra}). The same six stars selected for the CMD analysis in the main text are used here. The additional colour combinations provide an independent consistency check on the stellar population properties inferred for Fornax-7.

\section{The Bayesian interpretation of the metric $Q$}
\label{app:baysian}

\newcommand{\age}{\mbox{$\tau$}}
\newcommand{\FeH}{\mbox{[M/H]}}
\newcommand{\DM}{\mbox{$\mu$}}
\newcommand{\IMF}{\mbox{IMF}}
\newcommand{\mass}{\mbox{mass}}
\newcommand{\src}{\mbox{src}}

\newcommand{\NBS}{\mbox{NBS}}
\newcommand{\LF}{\mbox{LF}}
\newcommand{\Ld}{\mbox{Ld}}
\newcommand{\Cd}{\mbox{Cd}}
\newcommand{\CMD}{\mbox{CMD}}

An individual stochastic stellar-population model in our simulation-set is characterised by $M=\,$(\age, \FeH,\,\DM, \mass, \IMF) and by the seed $s$ of the random-number generation.  Under simplifying assumptions, and with $Q(M)$ defined in Eq.\,(\ref{eq1}) in Sect.\,\ref{sec:forward_modeling}, $~\mathrm{e}^{{-Q(M)}/{2}}$  can be interpreted as (proportional to) a posterior probability distribution of the model parameters that $M$ encapsulates. With $X$ standing for observables $X_1$ to $X_6$ in the numbered list in Sect.\,\ref{sec:forward_modeling}, 
the posterior probability of $M$ is written as
\begin{equation}
 P(M|X)  \ = \ 
  \frac{P(X | M)\,P(M)}{\sum_M P(X|M)\,P(M)}\,.
\label{eq:Bayes}
\end{equation}
For our given set of observations of Fornax-7, the denominator is a constant. Our priors $P(M)$ are given by the way we sampled age, metallicity, distance modulus, mass and the IMF when producing the set of simulations (cf. Sect.\,\ref{sec:forward_modeling}). For instance, the age grid is uniformly sampled on a log-scale, i.e. the implicitly assumed prior probability distribution of age itself is proportional to (age)$^{-1}$ between the adopted age boundaries. 
{\color{black} The prior mass distribution is only approximately flat. Indeed, the practical implementation we have adopted has a flat prior for the number of stars in Fornax-7 (on a linear scale between 1 and a number large enough to encompass all numbers of interest). The conversion of star number counts into mass has an average trend with age and a stochastic behaviour because the total number of stars in successful models is typically smaller than $10^3$.}

We implemented a 2-step approach for the evaluation of $P(X|M)$ by treating observable constraints $X_1$ to $X_3$, and $X_4$ to $X_6$, differently. To be
more specific, let $X_1$ (resp. $X_2, X_3$) stand for ``Constraint 1 (resp. 2, 3) is fulfilled''.
Let $X_4$ stand for ``the predicted colours of the BEST-6 stars in the CMD are (within a small interval of) those observed''; a similar definition holds for $X_5$ and $X_6$. Then
\begin{equation}
  P(X \, | \, M) \ = \ P( X_1,X_2,X_3 \text{~and~} X_4,X_5,X_6 \, | \, M) 
     \ = \ P(X_4,X_5,X_6 \, | \,  M, X_1,X_2,X_3 )\, P(X_1,X_2,X_3 \, | \, M)\,.
     \label{eq:app3-2}
\end{equation}
Factor $P(X_1,X_2,X_3 \, | \, M)$ is the probability $P_{\mathrm{occ}}$ of Sect.\,\ref{sec:forward_modeling}. Assuming $X_4$, $X_5$, $X_6$ are independent, 
$P(X_4,X_5,X_6 \, | \,  M, X_1,X_2,X_3 )$ can be rewritten as the product of 
the $P(X_i \, | \,  M, X_1,X_2,X_3 )$ with $i$ in $\{4,5,6\}$. If all errors are Gaussian and independent, these are proportional to $\mathrm{e}^{(-{{\chi^{2}_{i}}}/{2})}$. Their logarithm is illustrated in Fig.\,\ref{fig:f7-prob}. Finally,
\begin{equation}
- 2 \ln \left[ P(X \, | \, M) \right] \ = \  \sum_{i=4}^6 \chi^2_i - 2 \ln P_{\mathrm{occ}} \ = \ Q(M)\,, 
\label{eq:app3-metric}
\end{equation}
which is illustrated in Fig\,\ref{fig:f7-metric}.

\section{Forward-modelling results assuming a Chabrier IMF}
\label{app:chabrier}

\begin{figure}[h!]
    \centering
    \includegraphics[width=0.33\linewidth]{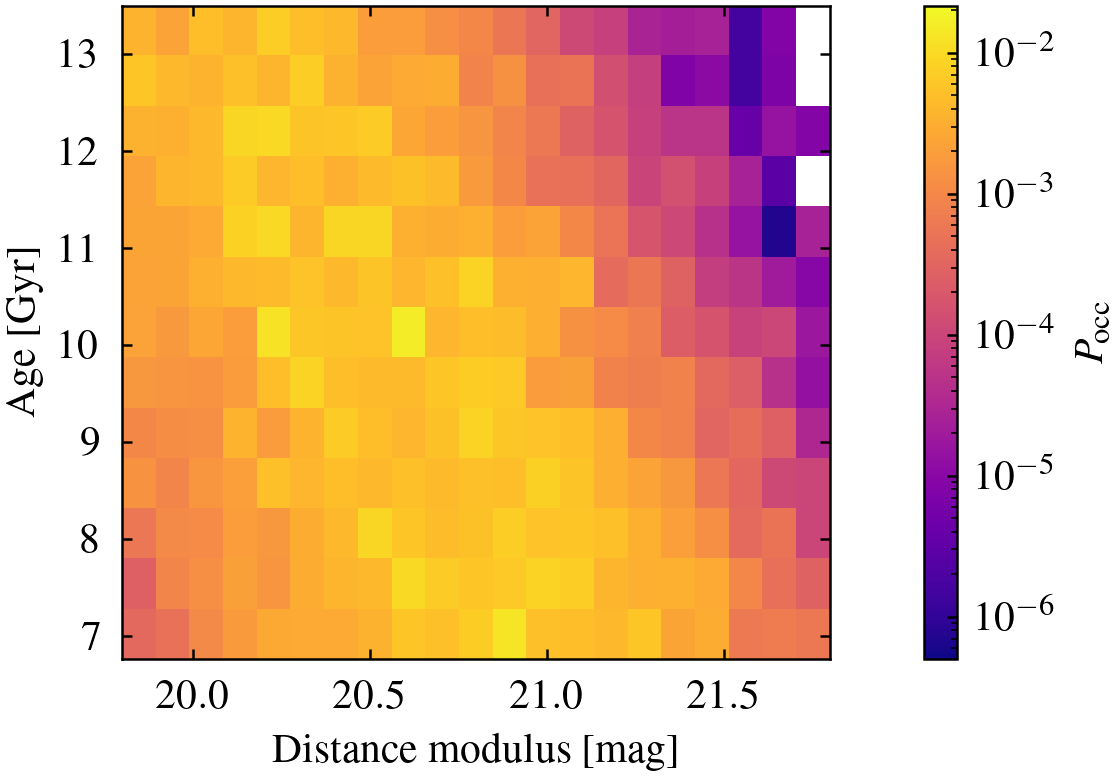}
    \caption{As Fig.\,\ref{fig:f7-prob} but assuming a Chabrier IMF.}
    \label{fig:f7-chabrier-prob}
\end{figure}

\begin{figure*}[h!]
    \centering
    \includegraphics[width=\linewidth]{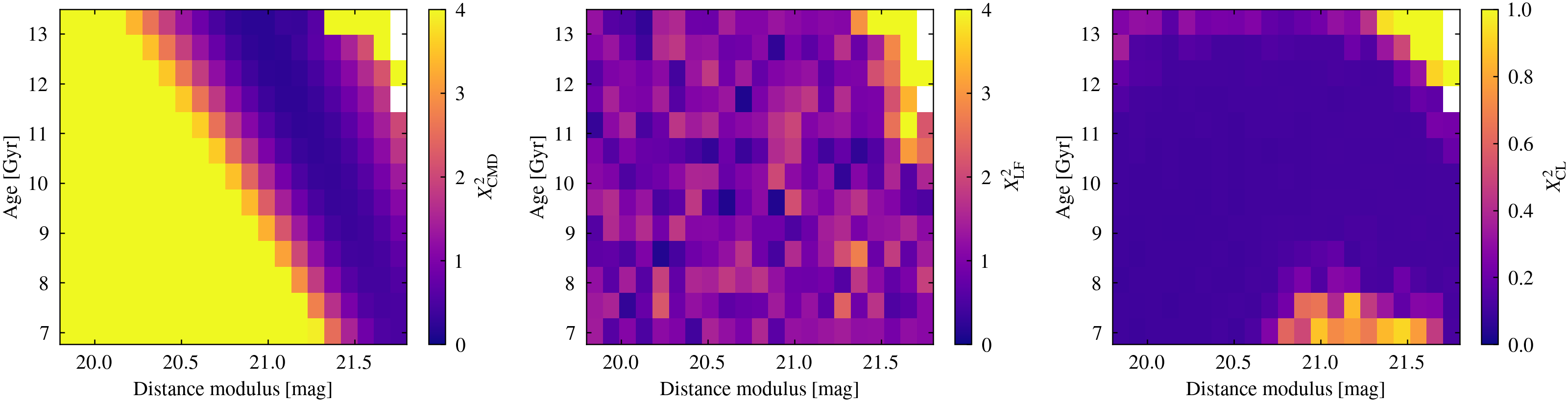}
    \caption{As Fig.\,\ref{fig:f7-chi2} but assuming a Chabrier IMF.}
    \label{fig:f7-chabrier-chi2}
\end{figure*}

\begin{figure}[h!]
    \centering
    \includegraphics[width=0.33\linewidth]{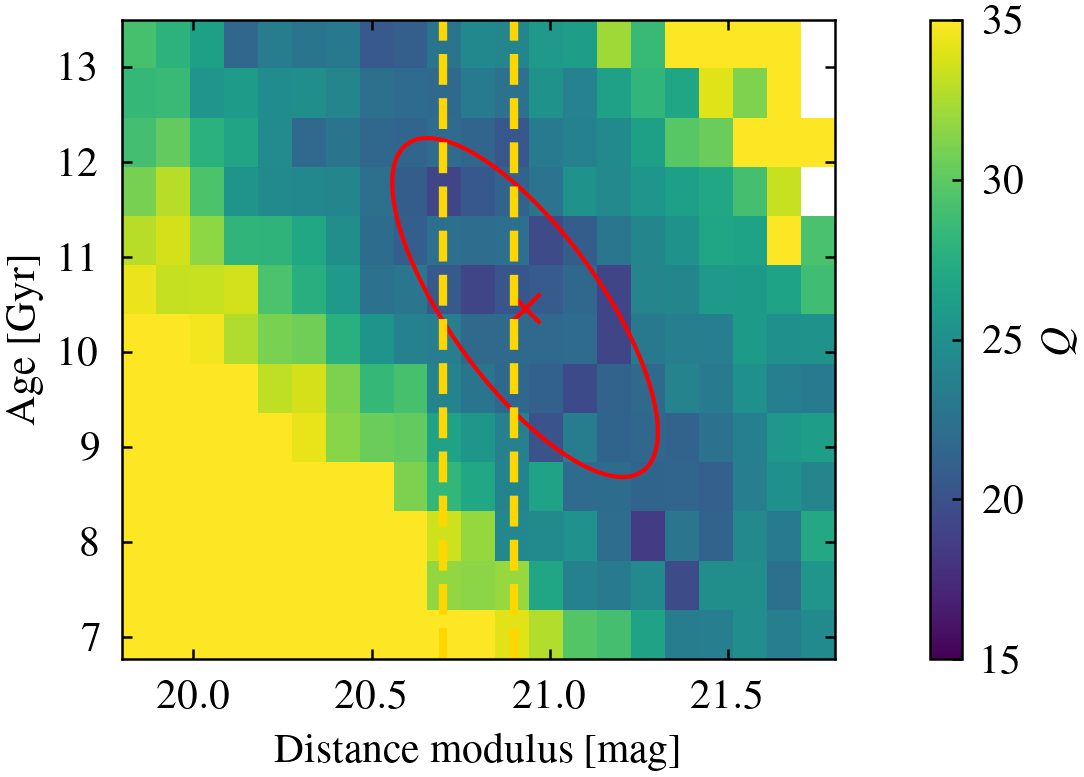}
    \hspace{2mm}
    \includegraphics[width=0.33\linewidth]{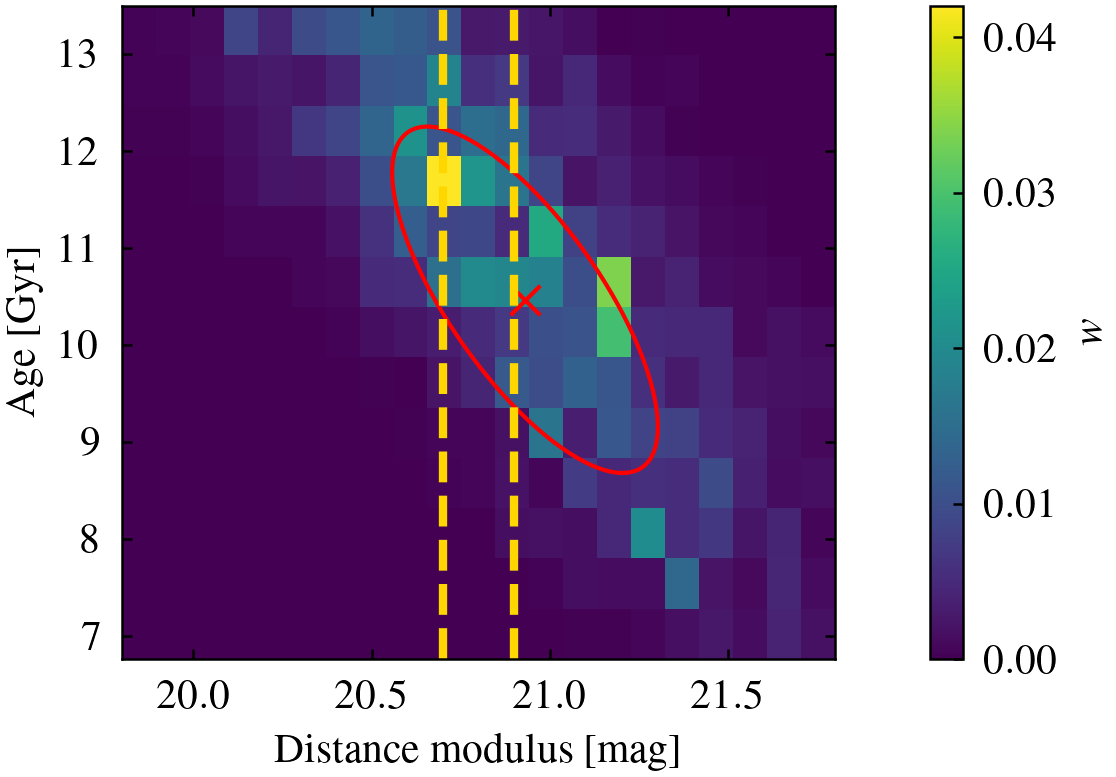}
    \hspace{2mm}
    \caption{As Fig.\,\ref{fig:f7-metric} but assuming a Chabrier IMF.}
    \label{fig:f7-chabrier-metric}
\end{figure}

Figures\,\ref{fig:f7-chabrier-prob}, \ref{fig:f7-chabrier-chi2}, and \ref{fig:f7-chabrier-metric} present the equivalent forward-modelling results obtained assuming a Chabrier IMF. The methodology is identical to that described in Sect.\,\ref{sec:forward_modeling}, with the only difference being the adopted stellar IMF used to generate the stochastic stellar populations. The resulting parameter distributions are highly consistent with those obtained using the Kroupa IMF. 

\section{Forward-modelling results for the relaxed metallicity constraints}
\label{app:relaxed}

\begin{figure}[h!]
    \centering
    \includegraphics[width=\linewidth]{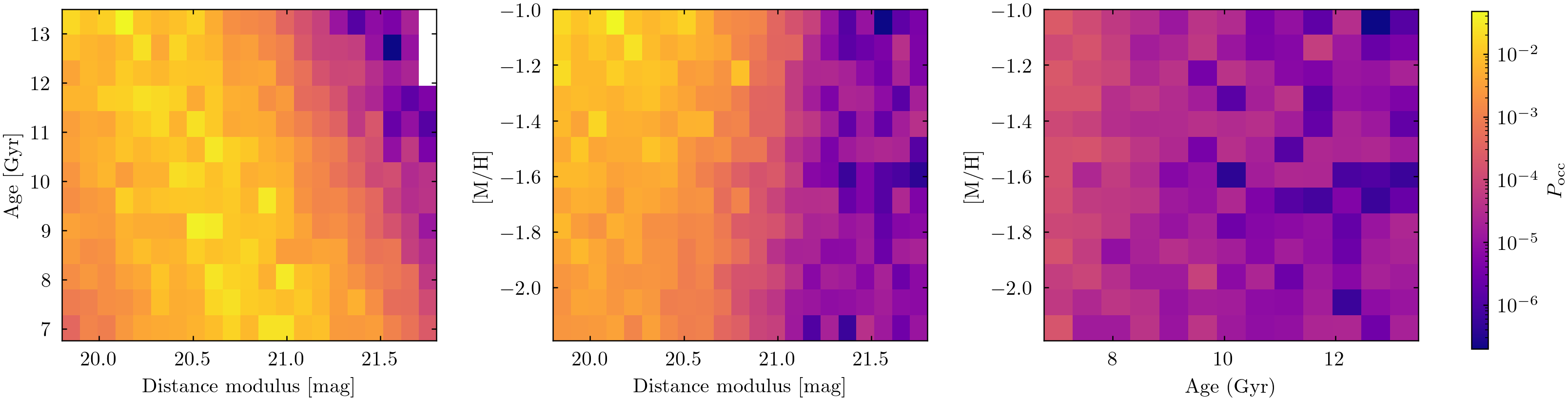}
    \caption{As Fig.\,\ref{fig:f7-prob} but for the relaxed metallicity constraints, and across distance modulus, age, and metallicity.}
    \label{fig:f7-kroupa-prob-relaxed}
\end{figure}

\begin{figure*}[h!]
    \centering
    \includegraphics[width=\linewidth]{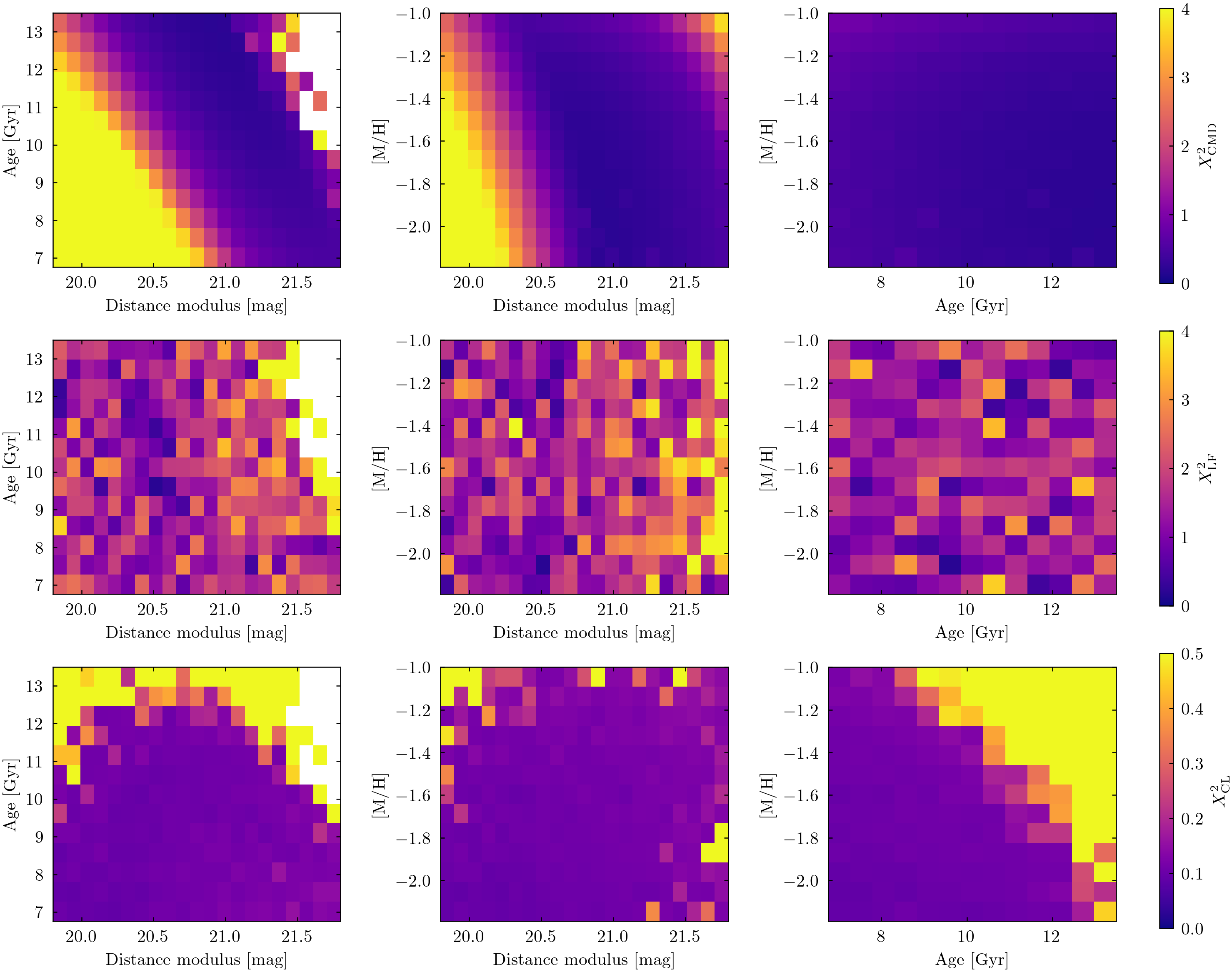}
    \caption{As Fig.\,\ref{fig:f7-chi2} but for the relaxed metallicity constraints, and across distance modulus, age, and metallicity.}
    \label{fig:f7-kroupa-chi2-relaxed}
\end{figure*}

\begin{figure}[h!]
    \centering

    \includegraphics[width=0.99\linewidth]{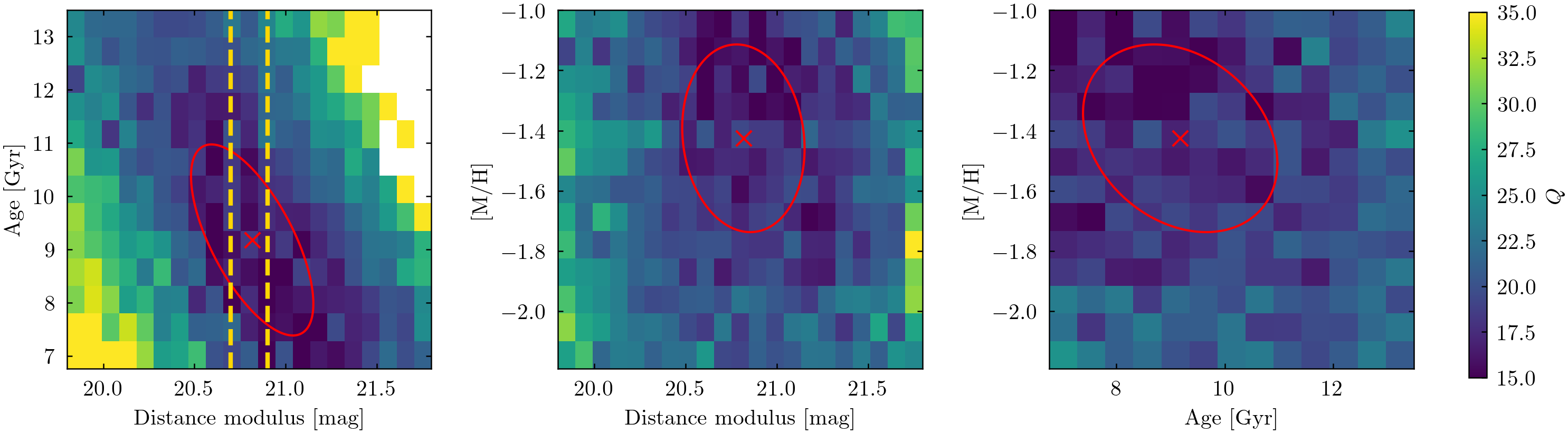}
    \hspace{2mm}
    \caption{As Fig.\,\ref{fig:f7-metric} but for the relaxed metallicity constraints, and across distance modulus, age, and metallicity.}
    \label{fig:f7-kroupa-metric-relaxed}
\end{figure}

Figures\,\ref{fig:f7-kroupa-prob-relaxed}, \ref{fig:f7-kroupa-chi2-relaxed} and \ref{fig:f7-kroupa-metric-relaxed} present the equivalent forward-modelling results obtained assuming a Kroupa IMF and relaxed metallicity constraints covering a broader range of metallicities. 

\end{appendix}

\label{LastPage}
\end{document}